\documentclass[aps,pra,twocolumn,superscriptaddress]{revtex4-1}

\usepackage[english]{babel}
\usepackage{amsmath}
\usepackage{amsthm}
\usepackage{amssymb}
\usepackage{mathtools}
\usepackage{ragged2e}
\usepackage{etoolbox}
\usepackage{booktabs}
\usepackage{color}
\usepackage{hyperref}
\usepackage{cases}
\usepackage{caption}
\usepackage{enumerate}
\usepackage{enumitem}
\usepackage{multirow}
\usepackage{slashed} 
\usepackage{feynmf}
\hypersetup{
  colorlinks   = true,  
  urlcolor     = blue,  
  linkcolor    = blue,  
  citecolor    = red    
}

\DeclareMathOperator{\Tr}{Tr}

\usepackage[font=small,labelfont=bf,justification=justified,format=plain]{caption} 
\usepackage{natbib}
\usepackage{pgfplots}
\pgfplotsset{compat=1.18}
\usepackage{subcaption}
\usepackage{url}
\newcommand{\ket}[1]{ | #1 \rangle }
\newcommand{\bra}[1]{ \langle #1 | }

\newcommand{\rom}[1]{\uppercase\expandafter{\romannumeral #1\relax}}

\usepackage{scalerel,stackengine}
\stackMath
\newcommand\reallywidehat[1]{%
\savestack{\tmpbox}{\stretchto{%
  \scaleto{%
    \scalerel*[\widthof{\ensuremath{#1}}]{\kern-.6pt\bigwedge\kern-.6pt}%
    {\rule[-\textheight/2]{1ex}{\textheight}}
  }{\textheight}%
}{0.5ex}}%
\stackon[1pt]{#1}{\tmpbox}%
}
\usepackage{comment}
\usepackage{float}

\usepackage{graphicx}
\usepackage{dcolumn}
\usepackage{bm}
\newcolumntype{M}{>{$}l<{$}}

\begin{document}

\title{Production of entangled x rays through nonlinear double Compton scattering }

\author{T. D. C. \surname{de Vos}}
\email{t.d.c.d.vos@tue.nl}
\affiliation{Department of Applied Physics and Science Education, Eindhoven University of Technology, P. O. Box 513, 5600 MB Eindhoven, The Netherlands}

\author{J. J. \surname{Postema}}
\affiliation{Department of Applied Physics and Science Education, Eindhoven University of Technology, P. O. Box 513, 5600 MB Eindhoven, The Netherlands}

\author{B. H. \surname{Schaap}}
\affiliation{Department of Applied Physics and Science Education, Eindhoven University of Technology, P. O. Box 513, 5600 MB Eindhoven, The Netherlands}
\affiliation{Department of Physics and Astronomy, UCLA, Los Angeles, California 90095, United States}

\author{A. Di \surname{Piazza}}
\affiliation{Department of Physics and Astronomy, University of Rochester, Rochester, New York 14627, United States}
\affiliation{Laboratory for Laser Energetics, University of Rochester, Rochester, New York 14623, United States}
\affiliation{Max Planck Institute for Nuclear Physics, Saupfercheckweg 1, D-69117 Heidelberg, Germany}

\author{O. J. \surname{Luiten}}
\affiliation{Department of Applied Physics and Science Education, Eindhoven University of Technology, P. O. Box 513, 5600 MB Eindhoven, The Netherlands}

\date{\today}

\begin{abstract}
An accessible tabletop source for the production of entangled x rays is crucial for the field of high-energy quantum optics. Here, we present a detailed analysis of the entanglement and polarization of the two photons emitted by an electron in an intense laser wave (nonlinear double-Compton scattering), by working within the framework of strong-field QED. By identifying contributions to the emission probability stemming from the electron being either on-shell or off-shell between the two-photon emissions, we show that the entangled photons are generated via the off-shell contribution for moderate beam energies, which are realizable on a tabletop setup. We propose an experiment to produce and isolate pairs of entangled x rays, through spectral filtering.
\end{abstract}

\maketitle

\section{Introduction}
Recently, strong interest has arisen from the field of quantum optics to stretch its boundaries to the x-ray spectral range, with interesting applications ranging from quantum lithography to x-ray cavities \cite{litho,Stopping_Narrow-Band_X-Ray,xray_applications1,xray_applications2}. Suggested sources of entangled x rays are free electron lasers (FELs), synchrotrons \cite{FEL_xrays,x-ray_beamsplitters1}, x-ray beam splitters \cite{x-ray_beamsplitters1,x-ray_beamsplitters2}, or Unruh radiation sources \cite{Unruh_radiation_R_Schutzhold_2008}. In this work we further build towards a tabletop entangled x-ray source by analyzing the production of entangled x rays through (nonlinear) double-Compton scattering, with moderate electron beam energies $\varepsilon\sim 10-100$ MeV. We note in passing that the relation between the Unruh effect and Compton scattering is under investigation \cite{Unruh_Classical_Larmor}.

The photon emission probability of nonlinear double-Compton scattering, referring to the emission of two photons in an intense laser field (i.e., $e^-+s\gamma_0\rightarrow e^- +\gamma_1+\gamma_2$, $s\in \mathbb{N}$), is a well researched topic \cite{HEITLER19341059,Daniel_Seipt,F.Mackenroth_Double_Compton_scattering,Victor_Double_Compton_Scattering,King_double_compton_scattering,Wang_Double_Compton_scattering,Torgrimsson_Double_Compton_Scattering}. It has been emphasized that the emission amplitude can be conveniently split into two contributions, depending on whether the electron between the two photon emissions is either on-shell ($\varepsilon^2-\mathbf{p}^2c^2 = m^2 c^4$) or off-shell ($\varepsilon^2-\mathbf{p}^2c^2 \neq m^2 c^4$). Since the electron dynamics in a plane-wave is intrinsically quasi-classical \cite{Ritus_1985}, the on-shell contribution can be interpreted as stemming from the electron propagating along classically allowed trajectories, which is clearly not the case for the off-shell contribution \cite{Daniel_Seipt}.

Unlike earlier works \cite{E.Lotstedt_PRA,E.Lotstedt_PRL,FEL_xrays,Unruh_radiation_R_Schutzhold_2006,Unruh_radiation_R_Schutzhold_2008}, we investigate the polarization, entanglement and probability of the photon emission process, allowing for a full classification of the entangled states, which is an essential aspect for experimental applications of an entangled x-ray source. For this we apply a finite laser pulse treatment, which is necessary to quantitatively talk about entanglement in our work. The energies required here to measure entangled x rays are significantly lower than those in Refs. \cite{E.Lotstedt_PRA,E.Lotstedt_PRL,FEL_xrays}, which renders the experimental observation of the entanglement in principle already feasible today. 

The structure of this paper is as follows. In Sec. \ref{Notation_conventions} we sum up the notations and conventions used throughout the paper. A brief introduction to the strong-field QED formulation is given in Sec. \ref{Formulation}. How to apply the generalized Stokes parameters and concurrence to this formulation is discussed in Sec. \ref{Polarization_Concurrence}. In Sec. \ref{Results} the entanglement, polarization and photon emission probability results of double-Compton scattering are presented. We find that for moderate beam parameters, the entangled x rays are produced by the off-shell channel. Also, we show that at the forbidden angle of Thomson scattering, where in a classical framework no radiation is emitted, a relative increase in the entangled photon pair production is found, compared to the unentangled contribution, which is in agreement with Refs. \cite{Unruh_radiation_R_Schutzhold_2006,Unruh_P_Chen_1999}. In Sec. \ref{Intuitive_Picture} we discuss a possible intuitive picture of the electron dynamics. In Sec. \ref{Experimental_Proposal} we propose a concrete method to measure entangled x rays with a tabletop setup. In Sec. \ref{Summary} the results are summarized.

\section{Notation and conventions}\label{Notation_conventions} 
Below, we use $a\cdot b \equiv a^\mu b_\mu$ as a Lorentzian inner product and employ natural units $c=\hbar=\epsilon_0=1$. We work in the light-cone coordinate system determined by the quantities: $k = \omega_0(1,\hat{\mathbf{n}}_0)$, $k_+ = \omega_0(1,-\hat{\mathbf{n}}_0)$, $\epsilon_1$, and $ \epsilon_2$, with $\{\epsilon_1,\epsilon_2\}$ being two four-vectors normalized to minus one and orthogonal to each other as well as to $k$ and to $k_+$. Note that $k_+$ is not a four-vector, however, here we write it in four-vector notation out of convenience. This allows for the decomposition of the space-time position $x^\mu$ as $x_- = x\cdot k/\sqrt{2} \omega_0$, $x_+ = x\cdot k_+/\sqrt{2} \omega_0 $ and $\mathbf{x}_\perp = -(x\cdot \epsilon_1,x\cdot \epsilon_2)$.  Furthermore, the $(+,-,-,-)$ signature is used for the Minkowski metric $\eta^{\mu \nu}$. For contraction with the gamma matrices, we use the slash notation: $\slashed{A} = \gamma^\mu A_\mu$. As such, we will be adopting $u_\sigma(p)$ and $v_\sigma (p)$ for the spinors of fermions and anti-fermions in the chiral/Weyl basis, respectively, normalized as $\overline{u}_\sigma(p) u_{\sigma'}(p) = -\overline{v}_\sigma(p) v_{\sigma'}(p) = 2m \delta_{\sigma \sigma'}$, with $m$ being the electron mass. Here, we are considering a finite plane-wave laser field only dependent on $\phi=\omega_0(t-\hat{\mathbf{n}}_0\cdot \mathbf{x})$, where $A_B(\phi)= \mathcal{A}(\phi)\epsilon+c.c.$ is the four-potential of the laser field with four-polarization $\epsilon$ and angular frequency $\omega_0$. This is an appropriate approximation if the electron beam waist is relatively small compared to the laser waist and fully classical in nature. The classical nonlinearity parameter is given by $a_0 = \sqrt{2} |e|  A_0/ m$ \cite{Ritus_1985,Di_Piazza_2012,Gonoskov_2022,Fedotov_2023}, where $e$ is the electron charge and $A_0 \equiv \max\{|\mathcal{A}|\}_\phi$. The $j$-th emitted photon wavevector is defined by: $q_j = \omega_j(1,\sin\theta_j\cos\varphi_j,\sin\theta_j\sin\varphi_j,\cos\theta_j)$, which can be normalized by: $n_j = q_j/\omega_j$, with $j\in \{1,2\}$. By indicating the initial four-momentum of the electron as $p_0 = m\gamma_0(1,\boldsymbol{\beta}_0)$, with $\gamma_0 = (1-\boldsymbol{\beta}_0^2)^{-1/2}$ being the Lorentz factor, the recoil parameter is given by $r = (p_0\cdot k)/m^2$. We will use $p_2$ to refer to the final four-momentum of the electron.

\section{Formulation}\label{Formulation} 
By working within the Furry picture in order to take into account the effects of the plane-wave exactly \cite{Furry_1951}, which is necessary for $a_0\gtrsim 1$ \cite{Di_Piazza_2012,Gonoskov_2022,Fedotov_2023}, the leading-order S-matrix element of the emission of two photons is given by $\mathcal{S}_2=\bra{p_2 q_1 q_2} S_2 \ket{p_0}$. Where $S_2 = T\left\{\left( -ie\int \,d^4 x  \overline{\psi} \slashed{A} \psi \right)^2\right\}/2$ is the $2$\textsuperscript{nd}-order scattering matrix, with $\psi(x)$ being the Dirac field to be expanded into Volkov states, i.e. the exact solutions of the Dirac equation in the presence of the plane-wave $A_B(\phi)$ \cite{Di_Piazza_2012,Gonoskov_2022,Fedotov_2023}, whereas $A(x)$ represents the quantized radiation field and $T$ denotes the time ordering operator. The scattering amplitude $\mathcal{M}_2$ will be defined by: $\mathcal{S}_2= (2\pi)^3  \delta^3\left(   p_{2}-p_{0}+q_{1}+q_{2} \right)_{- \perp}i \mathcal{M}_2 /\sqrt{2}\omega_0 $.   
\begin{figure}[t]
\centering
\includegraphics[scale=0.5]{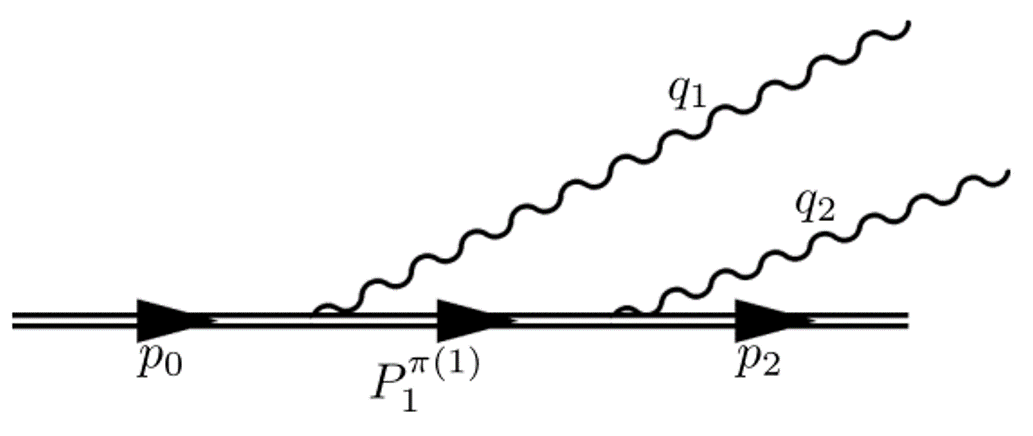}
\captionsetup{justification=Justified}
\caption{The leading-order ($\pi(1)$ permutation) Feynman diagram of nonlinear double-Compton scattering.}
\label{fig:Feynman}
\end{figure} 
The two-photon emission amplitude is then expressed as $\mathcal{M}_2 =  \mathcal{M}_2^{\pi(1)}+\mathcal{M}_2^{\pi(2)}$, $ \mathcal{M}_2^{\pi(1)/\pi(2)}$ being the $1/2$-permutation of the emitted photons. For the $\pi(1)/\pi(2)$ permutation the $q_{1/2}$ photon is emitted first.
The $\pi(1)$ permutation of this process can be expressed as a Feynman diagram as shown in Fig. \ref{fig:Feynman}, which depicts each vertex to be an effective channel of photon emission. The double arrows in Fig. \ref{fig:Feynman} represent the Volkov states and propagator, with $P_1 = p_1 -(p_1^2-m^2)/(2p_1\cdot k) k$ being the on-shell projection ($P_1^2=m^2$) of the internal electron four-momentum $p_1$, note that $P_0=p_0$ and $P_2=p_2$ because we assume that the initial and final electron four-momentum are on-shell.
Unlike in the linear case, the four-momentum of the internal electron in nonlinear double-Compton scattering can go on-shell. These on-shell electrons can propagate for a potentially macroscopic distance. As a result, the photons emitted through the on-shell channel turn out to be unentangled for $r\ll 1$. However, for high recoil the photon emission through this channel may become entangled due to the higher degree of inter-dependence of the two-photon emission events. Thus, it is of interest here to separate this on-shell channel from the remaining off-shell contribution, in order to gain more insight, see also Refs. \cite{Daniel_Seipt,F.Mackenroth_Double_Compton_scattering,Victor_Double_Compton_Scattering,King_double_compton_scattering,Wang_Double_Compton_scattering,Torgrimsson_Double_Compton_Scattering}. Since the on-shell contribution dominates for long laser pulses \cite{E.Lotstedt_PRA}, it is crucial for our case-study to do the calculations for a (short) finite laser pulse. The scattering amplitudes calculated through Feynman diagrams can be related to the differential two-photon emission probability (See Appendix \ref{N_Photon_Emission_Probability} and Ref. \cite{Daniel_Seipt}):
\begin{equation}
        \partial_{\omega,\Omega}^2 W = \frac{\omega_1 \omega_2}{16 (2 \pi)^6 (p_0 \cdot k) (p_2 \cdot k)} |\mathcal{M}_2|^2.
\end{equation}
This differential two-photon emission probability can conveniently be normalized by the laser frequency: $\hat{\partial}_{\omega,\Omega}^2 W= \omega_0^2 \partial_{\omega,\Omega}^2 W$.
We note that for a monochromatic (or long) pulse, due to energy-momentum conservation, there are resonance frequencies for $s$-photon absorption given by
\begin{equation}\label{eq:wres}
    \omega_2^{\rm res} = \frac{s k \cdot p_0 -q_{1} \cdot (s k+\overline{p}_0) }{n_2\cdot (sk+\overline{p}_0-q_{1}) },
\end{equation}
where we have adopted the notation of the effective electron momentum $\overline{p}_0 = p_0+m^2 a_0^2/(2 p \cdot k) k $, which manifests itself by the dressing of the electron in an infinite plane-wave background field. The Volkov Green's function (see also Ref. \cite{Ritus_1985}) can be separated into an on- and off-shell contribution, i.e. the on-, off-shell and total Volkov Green's functions in energy-momentum space are given by:
\begin{widetext}
\begin{equation}\label{eq:pulsedqqGon}
    \begin{split}
        i G_0^{\text{On}}(q,q') &= (2 \pi)^3 \delta^2(\mathbf{q}_\perp-\mathbf{q}'_\perp) \delta(k\cdot (q-q'))\frac{1}{4 |q\cdot k|} \int\,d \phi \,d \phi' \exp\left[iS^M_{q}(\phi)-i S^M_{q'}(\phi')\right]  \\ & \times  \left[1+e \frac{\slashed{k}  \slashed{A}_B(\phi)}{2 (q\cdot k)} \right]  \left[\slashed{q}'-\frac{q'^{2}-m^2}{2 q'\cdot k}\slashed{k}+m  \right] \left[1+e \frac{\slashed{A}_B(\phi') \slashed{k}  }{2 (q'\cdot k)} \right] ,
    \end{split}
\end{equation}

\begin{equation}\label{eq:pulsedqqGoff}
    \begin{split}
        i G_0^{\text{Off}}(q,q') &= (2 \pi)^3 \delta^2(\mathbf{q}_\perp-\mathbf{q}'_\perp) \delta(k\cdot (q-q'))\frac{1}{4 q\cdot k} \int\,d \phi \,d \phi' \exp\left[iS^M_{q}(\phi)-i S^M_{q'}(\phi')\right]    \left[1+e \frac{\slashed{k}  \slashed{A}_B(\phi)}{2 (q\cdot k)} \right]\\ & \times\left[\left(\slashed{q}'-\frac{q'^{2}-m^2}{2 q'\cdot k}\slashed{k}+m \right) \text{sgn}(\phi-\phi')+ i 2 \slashed{k}\delta(\phi-\phi') \right] \left[1+e \frac{\slashed{A}_B(\phi') \slashed{k}  }{2 (q'\cdot k)} \right],
    \end{split}
\end{equation}

\begin{equation}\label{eq:pulsedqqG}
    \begin{split}
    i G_0(q,q') &= (2 \pi)^3 \delta^2(\mathbf{q}_\perp-\mathbf{q}'_\perp) \delta(k\cdot (q-q'))\frac{1}{2 q\cdot k} \int\,d \phi \,d \phi' \exp\left[iS^M_{q}(\phi)-i S^M_{q'}(\phi')\right]    \left[1+e \frac{\slashed{k}  \slashed{A}_B(\phi)}{2 (q\cdot k)} \right]\\ & \times\left[\left(\slashed{q}'-\frac{q'^{2}-m^2}{2 q'\cdot k}\slashed{k}+m \right) \left(\Theta(q'\cdot k)\Theta(\phi-\phi')-\Theta(-q'\cdot k)\Theta(\phi'-\phi)\right)+ i \slashed{k}\delta(\phi-\phi') \right] \left[1+e \frac{\slashed{A}_B(\phi') \slashed{k}  }{2 (q'\cdot k)} \right]. 
\end{split}
\end{equation}
\end{widetext}

where $S^{M}_p(\phi) =  (p^2-m^2)/(2 p\cdot k) \phi -f_p(\phi)$, $f_p(\phi)= \int_{-\infty}^{\phi} d\phi' ( 2 e A_B\cdot p -e^2 A^2_B )/(2p\cdot k) $, $\Theta(\phi)$ is the Heaviside step function, $\text{sgn}(\phi)$ the sign function and $\delta(\phi)$ the Dirac delta function. We highlight that these results are in agreement with Ref. \cite{Off_on_Ilderton}. The total Green's function is, by definition, the summation of the on- and off-shell contributions, i.e. $G_0(q,q') = G_0^{\text{Off}}(q,q')+G_0^{\text{On}}(q,q')$. By solving the leading-order scattering amplitude $\mathcal{M}_2$, applying the appropriate Wick contractions and using the above total Green's function, we have \cite{F.Mackenroth_Double_Compton_scattering}
\begin{widetext}
\begin{equation}\label{eq:M2main}
\begin{split}
        i \mathcal{M}_2 &= (-ie)^2 \int \,d (\phi_2,\phi_{1}) \exp\left[i g_2^{\pi(1)}(\phi_2)+i g_1^{\pi(1)}(\phi_1)\right]  \overline{u}_{\sigma_2}(p_2) M_2^{\pi(1)}(\phi_2)  \left(   \left[\left(\slashed{P}_1^{\pi(1)}+m \right) \right. \right. \\& \times \left. \left. \left(\frac{\Theta(P_1^{\pi(1)}\cdot k)}{2 k\cdot P_1^{\pi(1)}}\Theta(\phi_{2}-\phi_1)+\frac{\Theta(-P_1^{\pi(1)}\cdot k)}{-2 k\cdot P_1^{\pi(1)}}\Theta(\phi_{1}-\phi_{2})\right)+ \frac{i \slashed{k}}{2 P_1^{\pi(1)}\cdot k}\delta(\phi_{2}-\phi_{1}) \right]  M_1^{\pi(1)}(\phi_1)\right) \\&  u_{\sigma_0}(p_0) + \pi(2) ,
\end{split}
\end{equation}
\end{widetext}
with $\pi(2)$ being a shorthand notation of $\mathcal{M}_2^{\pi(2)}$, $M_n^{\pi(1)}(\phi_n) = \slashed{\epsilon}_n^*-\frac{(\epsilon_n^* \cdot k) e^2 A_B^2(\phi_n)}{2(P_{n-1}^{\pi(1)}\cdot k)(P_n^{\pi(1)}\cdot k)}\slashed{k}+e \frac{\slashed{A}_B(\phi_n) \slashed{k} \slashed{\epsilon}_n^*  }{2 (P_n^{\pi(1)}\cdot k)} + e \frac{\slashed{\epsilon}_n^* \slashed{k}  \slashed{A}_B(\phi_n)}{2 (P_{n-1}^{\pi(1)}\cdot k)}$ is the $n$-th photon emission matrix and \\
$  g_{n}^{\pi(j)}(\phi_n) =\kappa_{n}^{\pi(j)}\phi_n+f_{P_n^{\pi(j)}}(\phi_n)-f_{P_{n-1}^{\pi(j)}}(\phi_n)$
is the $n$-th phase element. Additionally, the quantities $\kappa_2^{\pi(1)} = p_2\cdot q_2/(p_0-q_1)\cdot k$ and $\kappa_1^{\pi(1)} = p_0\cdot q_1/(p_0-q_1)\cdot k $ are the effective photon absorption parameters. The $\pi(2)$ permutation of the above parameters is trivially obtained by switching the photon four-momentum vectors $q_1,q_2$ and the polarization vectors $\epsilon_1,\epsilon_2$. The separation of the scattering amplitude into the two contributions where the intermediate electron propagation between photon emission events is either on-shell or off-shell is obtained by using the on- and off-shell Green's functions instead of the total Green's function. We will refer to these two separations as the on- and off-shell contributions of the scattering amplitude, respectively. It is straightforward to show that this translates to separating the Heaviside step functions into a sign function and a constant, e.g. $\Theta(\phi_2-\phi_1) = (\text{sgn}(\phi_2-\phi_1)+1)/2$, which corresponds to the off- and on-shell contribution of the scattering amplitude, respectively. The delta function part $\delta(\phi_2-\phi_1)$ in Eq. (\ref{eq:M2main}) only contributes to the off-shell component of the scattering amplitude. Additionally, we will use the on- and off-shell scattering amplitudes to calculate the on- and off-shell contributions of the photon emission, entanglement, and polarization. More details on the on- and off-shell splitting of the scattering amplitude can be found in Refs. \cite{Daniel_Seipt,Seipt_thesis}. In Appendixes \ref{N_Photon_Emission_Amplitude} and \ref{Double_Compton_Scattering_Amplitude_Analysis} we treat the general $N$-photon emission scattering amplitude, and conduct an analysis of the double-Compton scattering amplitude, respectively.

\section{Polarization and Concurrence}\label{Polarization_Concurrence} 
We use the generalized (two-qubit) Stokes parameters as a measure of polarization, as discussed in earlier works \cite{Stokes_1,Stokes_2}. To this end we define a density matrix of the form $\rho_{\beta_1 \beta_2 }^{\alpha_1 \alpha_2 } =   \mathcal{M}_{\alpha_1 \alpha_2 }  \mathcal{M}_{\beta_1 \beta_2}^*$, where $\{\alpha_{1},\alpha_2, \beta_{1},\beta_2\} = \{0,1\}^{\otimes 4}$ is the polarization basis of the emitted photons, averaged/summed over the initial/final electron spin. Here we dropped the $2$ subscript of $\mathcal{M}_2$, since it is clear by the polarization basis that we are considering two-photon emission. Note that $\rho$ can always be mapped onto a $4 \times 4$ matrix. We project the photon polarization on the polarization plane of the laser field (see Appendixes \ref{Entanglement_of_a_Two_Qubit_System} and \ref{Ap.Polarization_and_Stokes_Parameters} for further details), such that the polarization basis is $\ket{L} = (\ket{H}+i\ket{V})/\sqrt{2}$ and $\ket{R} = (\ket{H}-i\ket{V})/\sqrt{2}$ (with $\langle H | V \rangle=0$), which are the left- and right-handed polarization states, respectively. Moreover, $\ket{H}$ is polarized along the same direction as the laser polarization. Then, after normalizing the density matrix $\hat{\rho}= \rho/\Tr\{\rho\}$, the generalized Stokes parameters for two photons are:
\begin{equation}
    \hat{s}_{l_1 l_2} =  \Tr\left\{\hat{\rho}  (S_{l_1} \otimes S_{l_2}) \right\},
\end{equation}
where $S_l$ is the $l$-th Stokes parameter operator, with $\{l_1,l_2\} \in \{0,1,2,3\}^{\otimes 2}$. For the Stokes parameter operators, $S_0=\ket{H}\bra{H}+\ket{V}\bra{V}$ is the identity operator, $S_1=\ket{H}\bra{H}-\ket{V}\bra{V}$ is the degree of linear polarization, $S_2=\ket{D_+}\bra{D_+}-\ket{D_-}\bra{D_-}$ the degree of diagonal polarization defined by $\ket{D_\pm} = \frac{1}{\sqrt{2}}(\ket{H}\pm \ket{V})$, and $S_3=\ket{L}\bra{L}-\ket{R}\bra{R}$ the degree of circular polarization, which in the $\{L,R\}$ basis yields: $S^{\{L,R\}}_l= \sigma_l$. Where $\sigma_l$ is the $l$-th Pauli matrix. For the two-qubit Stokes parameters $\hat{s}_{11}$ is the (generalized) degree of linear polarization, $\hat{s}_{22}$ the degree of diagonal polarization and $\hat{s}_{33}$ the degree of circular polarization, where the polarization states are (un)equal for $\hat{s}_{ll}=+1(-1)$. For example, the state $\ket{HH}$ has $\hat{s}_{11}=1$ and $\hat{s}_{22}=\hat{s}_{33}=0$. A Stokes parameter of the form $\hat{s}_{12}$ should be interpreted as a measure to what degree photon $1$ is linearly polarized and photon $2$ diagonally polarized. For entangled states it is possible to have (minus) unity for multiple Stokes parameters, e.g. $\ket{\psi} = (\ket{HV}-\ket{VH})/\sqrt{2}= i (\ket{LR}-\ket{RL})/\sqrt{2} $ has $\hat{s}_{11}=\hat{s}_{33}=-1$. This is a direct consequence of the entangled nature of the state and the mapping between the different bases. This means that multi-particle states can simultaneously be entangled in both circularly polarized and linearly polarized bases. 

The entanglement for a two-photon system can be expressed via the concurrence. The concurrence is calculated by finding the eigenvalues $\lambda_n$ of $\hat{\rho} \tilde{\rho} $, with $\tilde{\rho} = \sigma_2^{\otimes 2} \hat{\rho}^* \sigma_2^{\otimes 2}$ (in the original $\{0,1\}^{\otimes 4}$ basis) being the spin flip density matrix, and sorting them in descending order, by evaluation of \cite{Concurrence}: 
\begin{equation}\label{eq:concurrence1}
    C(\hat{\rho}) = \max(0,\sqrt{\lambda_1}-\sqrt{\lambda_2}-\sqrt{\lambda_3}-\sqrt{\lambda_4}).
\end{equation}
The concurrence $C\in [0,1]$ is unity for a fully entangled state, e.g. a Bell state, and zero for an unentangled state. We would like to point out that states with a high concurrence, i.e. $C\approx 1$, are also highly uncoupled from the final electron spin, due to the nature of tracing out the final electron spin from the density matrix.
\begin{figure}[t]
    \centering
    \includegraphics[scale=0.70]{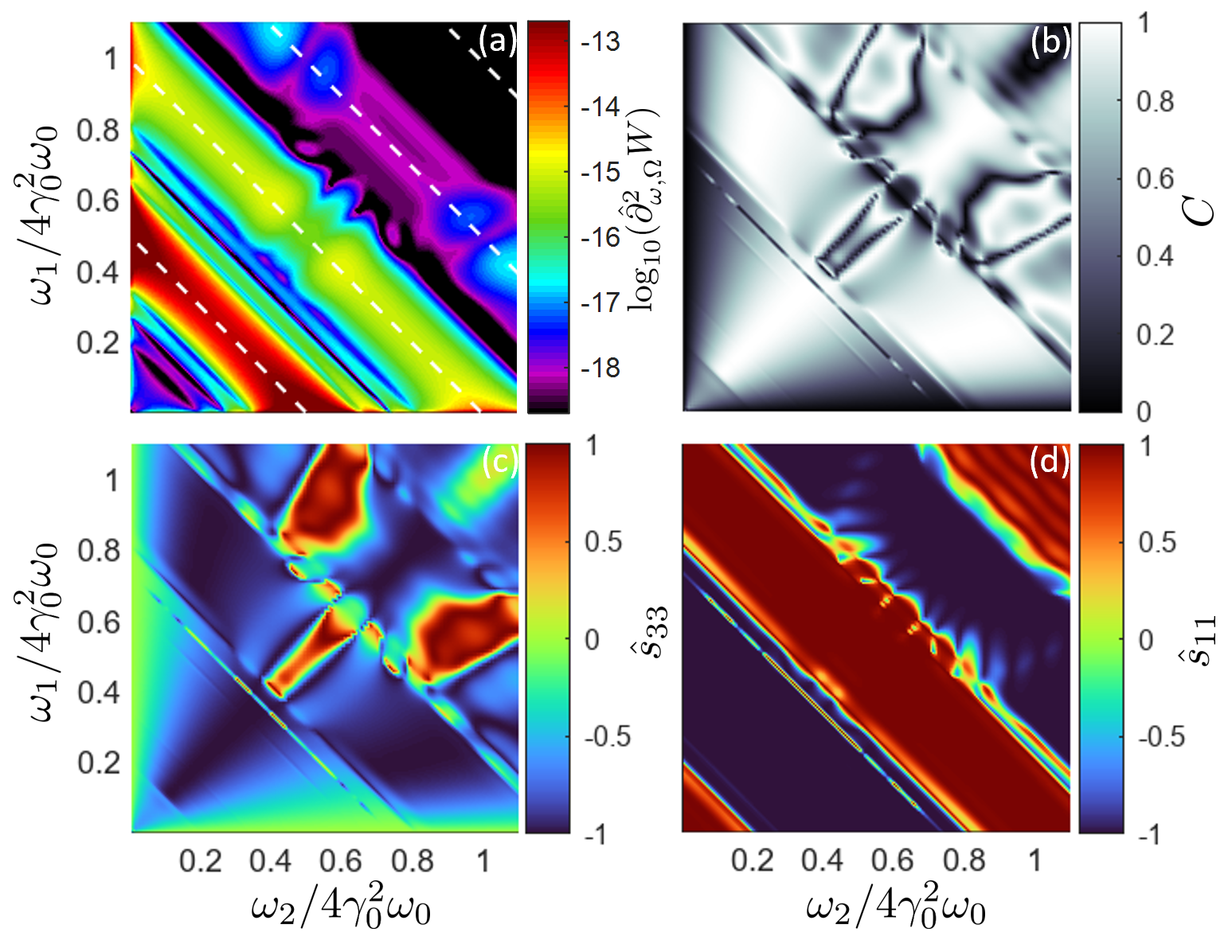}
    \captionsetup{justification=Justified}
    \caption{The off-shell spectral distribution of double-Compton scattering for an electron in a head-on scattering geometry with $a_0=0.1$, $\gamma_0 \approx 70.7$ ($|\boldsymbol{\beta}_0| = 0.9999$), $\omega_0/m=10^{-5}$, $\Delta \phi = 40$, $\theta_1=\theta_2=1/\gamma_0$, $\varphi_1 = \pi/2$ and $\varphi_2 = 3\pi/2$. Here \textbf{(a)} is the normalized differential photon emission probability $\log_{10}(\hat{\partial}_{\omega, \Omega}^2 W)$, with the white dashed lines indicating the $s=\{1,2,3,4\}$ photon emission resonance frequencies as in Eq. (\ref{eq:wres}), \textbf{(b)} the concurrence, \textbf{(c)} the generalized degree of circular polarization $\hat{s}_{33}$, \textbf{(d)} the generalized degree of linear polarization $\hat{s}_{11}$. The photon frequencies $\{\omega_1,\omega_2\}$ are scaled with $4 \gamma_0^2 \omega_0$, which is approximately the double-Doppler shift of the laser frequency for on-axis backward scattering of Thomson scattering.   }
    \label{fig:a0_01}
\end{figure}

\begin{figure}[b]
    \centering
    \includegraphics[scale=0.70]{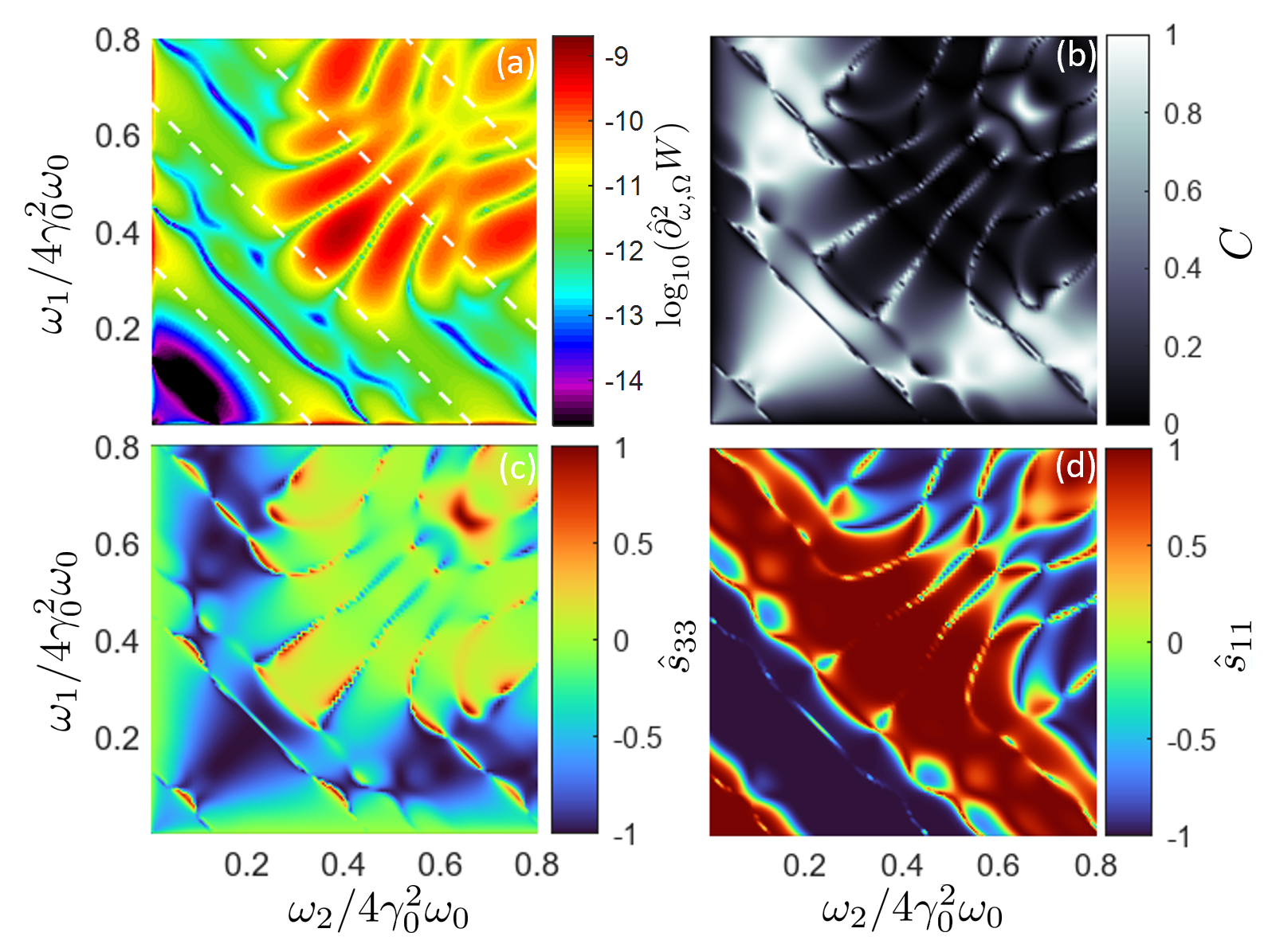}
    \captionsetup{justification=Justified}
    \caption{The off-shell spectral distribution of double-Compton scattering for an electron in a head-on scattering geometry with $a_0=1$, $\gamma_0 \approx 70.7$ ($|\boldsymbol{\beta}_0| = 0.9999$), $\omega_0/m=10^{-5}$, $\Delta \phi = 40$, $\theta_1=\theta_2=1/\gamma_0$, $\varphi_1 = \pi/2$ and $\varphi_2 = 3\pi/2$. Here, the layout and graphs are the same as in Fig. (\ref{fig:a0_01}). }
    \label{fig:a0_1}
\end{figure} 

\section{Results Double Compton Scattering}\label{Results}
Below, we consider the background laser field to be described by $\mathcal{A}(\phi) = A_0 \cos^2(\pi \phi/\Delta\phi) (\Theta(\phi+\Delta\phi/2)-\Theta(\phi-\Delta\phi/2)) \exp[-i\phi] $, so that the laser pulse is exactly $\Delta \phi$ long and has a full-width-half-maximum of $\Delta\phi/2$. Furthermore, we have a laser wave propagating along the $\hat{z}$ axis [$\mathbf{\hat{n}}_0=(0,0,-1)$] and polarized along the $\hat{x}$ axis [$\epsilon = (0,1,0,0)$]. We will only consider head-on scattering, hence the electron is traveling along the $\hat{z}$ axis [$\boldsymbol{\beta}_0=(0,0,\beta_0)$]. The on-shell contribution of double-Compton scattering has the same polarization as one would expect for two single-Compton scattering events for $r \ll 1$, which is shortly discussed in Appendix \ref{Polarization_of_Compton_Scattering} (for more information on single-Compton scattering, see Refs. \cite{cs1,cs2,cs4,cs5,cs6,cs6,cs8,Victor_Double_Compton_Scattering}). For double-Compton scattering we will focus on two specific orientations of the azimuthal angle of photon emission: $\{\varphi_1,\varphi_2\}=\{\pi/2,3\pi/2\}$ and $\{\varphi_1,\varphi_2\}=\{0,\pi\}$, that is orthogonal and parallel, respectively, to the laser wave polarization direction $\hat{x}$. For the first orientation we have the polar angles $\theta_1=\theta_2=1/\gamma_0$, and for the second we have $\theta_1=\theta_2=\sqrt{1+a_0^2}/\gamma_0$. 

\begin{figure}[t]
    \centering
    \includegraphics[scale=0.75]{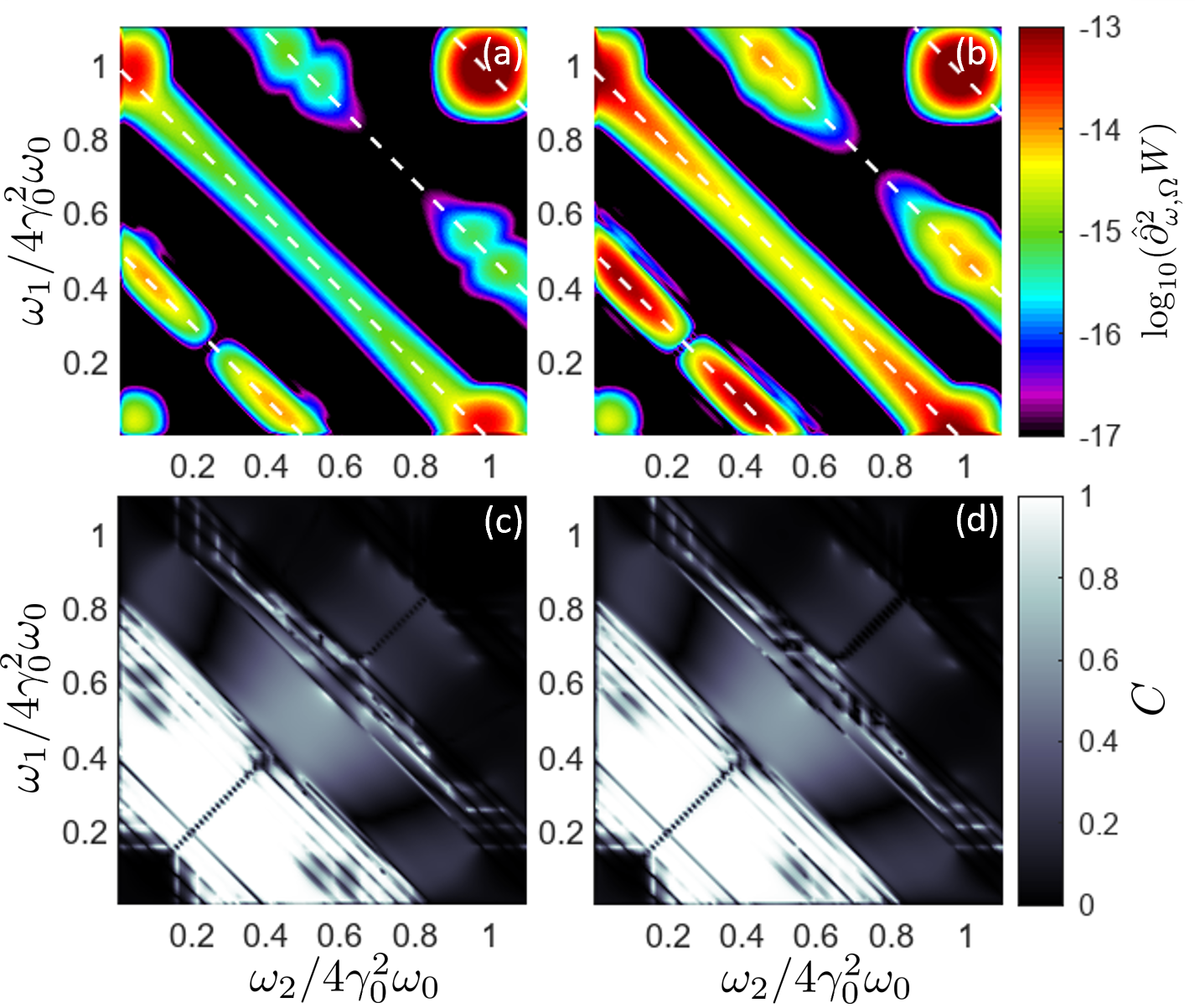}
    \captionsetup{justification=Justified}
    \caption{The total spectral distribution of double-Compton scattering for an electron in a head-on scattering geometry with $a_0=0.1$, $\omega_0/m=10^{-5}$, $\Delta \phi = 60$, $\theta_1=\theta_2=\sqrt{1+a_0^2}/\gamma_0 $, $\varphi_1 = 0$ and $\varphi_2 = \pi$. Here, \textbf{(a)} and \textbf{(c)} depicts the normalized differential photon emission probability $\log_{10}(\hat{\partial}_{\omega, \Omega}^2 W)$ and concurrence for electron energy $\gamma_0=70.7$, respectively. Where the white dashed lines indicate the $s=\{1,2,3,4\}$ photon emission resonance frequencies as in Eq. (\ref{eq:wres}). \textbf{(b)} and \textbf{(d)} are the same graphs but for electron energy $\gamma_0=200$.}
    \label{fig:deadangle}
\end{figure} 

Figures \ref{fig:a0_01} and \ref{fig:a0_1}, show the off-shell normalized differential photon emission (see Appendix \ref{Double_Compton_Scattering_Amplitude_Analysis} for the total and on-shell results), polarization and entanglement (concurrence) of photons emitted in opposite direction on the $\hat{y}$ axis, for $a_0=0.1$ and $a_0=1$, respectively. As shown in Fig. \ref{fig:a0_01} the degree of circular polarization is highly correlated with the entanglement at this angle. Focusing on Fig. \ref{fig:a0_01}\textcolor{blue}{.c} we see that at certain frequencies the polarization switches from $\hat{s}_{33}=-1$ to $\hat{s}_{33}=1$. This switch can be understood as a transitional state, which is mediated by the ponderomotive interaction. This effect can also be observed in Fig. \ref{fig:a0_1}. Additionally, at the frequencies where this ponderomotive broadening effect is most prominent, the entanglement decreases the most for $a_0=1$ compared to $a_0=0.1$. Note that despite this effect, still more entangled photons are being emitted for $a_0=1$. However, it is more difficult to separate these entangled photons from the unentangled photons, since due to ponderomotive broadening, the off- and on-shell contributions have a high degree of spectral overlap, rendering spectral filtering less applicable. The biggest disadvantage when limiting ourselves to moderate beam energies is that the off-shell (and therefore entangled) contribution is relatively small compared to the on-shell contribution, e.g. for the orientation and parameters discussed above one has a five orders of magnitude difference in the photon emission probability, making filtering techniques essential. However, there is an orientation for which the off-shell contribution is relatively enhanced, the $\{\varphi_1,\varphi_2\}=\{0,\pi\}$ orientation.   

Figure \ref{fig:deadangle} shows the normalized differential photon emission probability and concurrence (for $a_0=0.1$) for this orientation. In classical theory, i.e. Thomson scattering, one expects hardly any radiation at this angle since it is approximately parallel to the electron acceleration, for the considered beam energies. However, due to electron recoil it is possible to emit radiation at this angle via Compton scattering (see also Ref. \cite{F.Mackenroth_Double_Compton_scattering}). This does suppress the on-shell contribution significantly allowing for a relatively larger contribution of entangled x-ray radiation. We analyze two different initial electron Lorentz factors $\gamma_0 = 70.7$ ($\varepsilon \approx 36$ MeV) and $\gamma_0=200$ ($\varepsilon \approx 102$ MeV). For $\gamma_0 = 200$ we see that the entangled radiation is of the same order of magnitude as the unentangled radiation. We stress that, although $r\ll 1$ and $a_0 \in [0.1,1]$, a full quantum calculation is necessary as the emission of two photons and their entanglement have no classical interpretation.   

\section{Intuitive Picture}\label{Intuitive_Picture}

In Figs. \ref{fig:a0_01} and \ref{fig:a0_1} one observes a high degree of circular polarization, which is interesting since classically one does not expect any circularly polarized photons to be emitted in a linearly polarized laser pulse. Here we provide an intuitive picture explaining how an electron can emit circularly polarized photons, in a linearly polarized laser field.        

\onecolumngrid

\begin{figure}[H]
\captionsetup{justification=Justified}
\centering{
\includegraphics[width=0.95\textwidth]{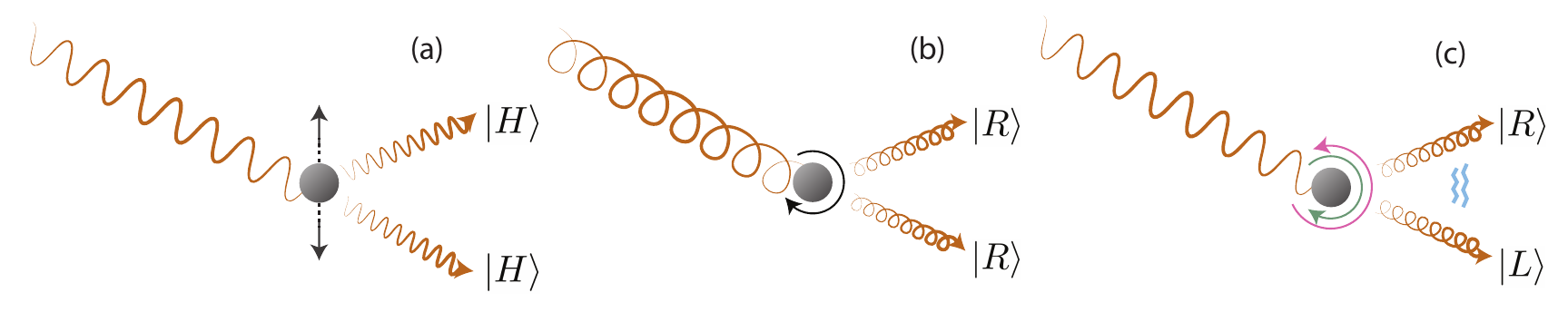}} 
\caption{Different two-photon emission processes (in the co-moving frame) that an electron in an external pulsed laser field can undergo. \textbf{(a)} An electron in a linearly polarized laser field emits linearly polarized photons. \textbf{(b)} Circularly polarized light will allow electrons to emit photons with the same polarization. \textbf{(c)} The production of an entangled photon pair in a linearly polarized laser field can be interpreted as an electron in a superposition of left- and right-handed circular paths. The emitted photons are entangled in the $\ket{L}, \ket{R}$-basis.}
\label{fig:photons}
\end{figure}

\twocolumngrid 

\noindent Let us consider such a laser field and compare the classical electron trajectory in this field (linear path) with the average of the classical paths in a left- and right-circularly polarized laser field. The difference between the linear path and the superposition of the left- and right-handed paths of the electron, i.e. $\Delta \mathbf{x}$, scales in the co-moving frame like: $|\Delta \mathbf{x}| \propto a_0^2$. For relatively low laser intensities such that $a_0 < 1$, the path difference $\Delta \mathbf{x}$ becomes relatively small, hence the off-shell electrons can be interpreted to be in a superposition of two circular paths, as illustrated in Fig. \ref{fig:photons}. Classically, an electron in a linearly polarized laser field will emit linearly polarized radiation through Thomson scattering \cite{Petrillo}. However, if two photons are emitted by the same electron, then the electron can emit left- and right-handed polarized photons, with the sum of the photons' energies equal to the energy of a single-Compton scattering event, or a higher harmonic thereof. These photons are mostly radiated via the off-shell contribution (as discussed above) and therefore, since the two emissions temporally occur very close to each other (in the lab-frame), their polarization states are highly correlated. This also allows us to intuitively understand why at the angle where the electron does not radiate classically, see Fig. \ref{fig:deadangle}, the off-shell contribution is less suppressed than the on-shell contribution. This can be explained by the understanding that the circular electron trajectories are in principle allowed to radiate at this angle.

\section{Experimental proposal}\label{Experimental_Proposal} 

Making use of x-ray polarizers and placing the x-ray polarizers \cite{Xray_polarimetry} perpendicular to the polarization direction (for the fundamental, i.e. $s=1$) as predicted by single-Compton scattering, would make it possible to very efficiently filter out the on-shell double-Compton scattering and single-Compton scattering contributions. However, we do note that this is a destructive filtering method. Alternatively, spectral filtering would allow for non-destructive filtering of the entangled x rays. For the experimental setup, $\gamma_0\approx 70$, $\omega_0\approx 1.6 $ eV, and a short laser pulse of intensity $I \approx 10^{17}$ W/cm$^2$ ($a_0\approx 0.1$) would suffice for creating sufficiently large flux of entangled x-ray photons for practical experiments. These parameters are already achieved at inverse Compton scattering setups \cite{ICS_Graves,ICS-Gunther} and entangled photons with energies $\approx 8  \times s $ keV would be emitted. However, higher electron energies would yield more entangled x rays, as shown in Fig. \ref{fig:deadangle} for $\gamma_0 = 200$. To get insight in the ratios of the off- and on-shell contributions $\mathcal{R}= \max\{\partial^2_{\omega, \Omega}W_{\text{Off}}\}_{\omega_{1/2}}/\max\{\partial^2_{\omega, \Omega}W_{\text{On}}\}_{\omega_{1/2}}$ for different electron energies, see Tab. \ref{tab:ratio_off_on}. The ratio $\mathcal{R}$ at $\varphi_1=0$ is several orders of magnitude higher than at $\varphi_1=\pi/2$. 
\begin{table}[H]
    \centering
    \caption{The order of magnitude of $\mathcal{R}$ for different electron energies. Here, $\theta_1=\theta_2=1/\gamma_0$, $\varphi_2-\varphi_1=\pi$, $\omega_0 = 1.6$ eV, $\Delta \phi=40$ and $a_0=0.1$.}
    \begin{tabular}{|c|c|c|c|c|}\hline
        $\gamma_0$ & $10^1$ & $10^2$ & $10^3$ & $10^4$  \\ \hline
       $\mathcal{O}\{\mathcal{R}\}(\varphi_1=\pi/2)$ & $10^{-7}$  & $10^{-5}$ & $10^{-3}$ & $10^{-1}$  \\ \hline
       $\mathcal{O}\{\mathcal{R}\}(\varphi_1=0)$ & $10^{-4}$  & $10^{-2}$ & $10^0$ & $10^2$  \\ \hline
    \end{tabular}
    \label{tab:ratio_off_on}
\end{table} 

\noindent This is due to the suppression of the on-shell contribution at the classically forbidden angle (that is, $\varphi_1=0$). Ideally one should take $\theta_1=\theta_2=\sqrt{1+a_0^2}/\gamma_0 $ for $\varphi_1=0$, to get the largest relative off-shell contribution, compared to the on-shell contribution. Note that for $\gamma_0 \gg 10^4$ we have $r > 0.1$, which would give rise to polarization mixing due to the high recoil, rendering polarization filtering less efficient.

\section{Summary}\label{Summary} 
In this article we build towards realizing a tabletop entangled x-ray source, by providing a theoretical framework to fully classify and quantify the entangled photon emission, via double-Compton scattering. We have shown that with current state-of-the-art tabletop experimental setups, it is possible to generate a significantly large flux of entangled x-ray photons, and isolate them from the unentangled photons by spectral filtering. 

\section{Acknowledgements}\label{Acknowledgements} 
We thank Coen Sweers and Simona Borrelli for the valuable discussions. This publication is part of the Industrial Partnership Program ColdLight, which is financed by the Dutch Research Council (NWO) and the Dutch company Acctec B.V.

\appendix

\section{$N$-photon emission probability}\label{N_Photon_Emission_Probability}
Here we calculate the photon emission probability as a function of the scattering matrix elements. For completeness, we consider that $N$ photons are emitted throughout the scattering process, such that $p_N$ is the final momentum. To calculate the differential photon emission probability for a finite plane-wave (FPW) we will take a scattering amplitude $\mathcal{M}_N$ as a given, defined by:
\begin{equation}\label{eq:Mdef}
    \mathcal{S}_N  = (2\pi)^3  \delta^3\left(   p_{2}-p_{0}+q_{1}^N \right)_{- \perp} i \mathcal{M}_N/\sqrt{2}\omega_0,
\end{equation}
where we have adopted the notation $q_1^n = \sum_{j=1}^n q_j$.
This already shows that for the photon emission of FPWs, we should consider the light-front phase-space to cancel out the light-front volumes. In the Furry picture the electron-photon scattering is considered as a decay mediated by the background field $A_B$.
Therefore, we use the standard definition of the decay probability, given by \cite{Schwartz}:
\begin{widetext}
\begin{equation}\label{eq:dW_N}
    d W = \frac{ |\mathcal{S}_N|^2}{\langle p_0 | p_0 \rangle \langle p_N q_1 \cdots q_N | p_N q_1 \cdots q_N \rangle} V_{\rm LF }\frac{d^3 p_N}{(2 \pi)^3}\prod_{n=1}^N V \frac{d^3 \mathbf{q}_n}{(2 \pi)^3},
\end{equation}
\end{widetext}
where $d^3 p_N \equiv dp_{N-} d^2\mathbf{p}_{N \perp}$ and $d^3 \mathbf{q}_n = d q_{n 1} d q_{n 2} d q_{n 3} $. $V_{\rm LF}$ and $V$ are the light-front volume and the Cartesian volume, respectively:  
\begin{align*}
    V_{\rm LF} &\equiv \int \,d^3 (x_-,\mathbf{x}_\perp) &  V  &\equiv \int \,d^3\mathbf{x},
\end{align*}
Taking ${p_0,~p_N}$ in the light-front basis and $q_n$ in the Cartesian basis, we have:
\begin{align*}
    \langle p_0 | p_0 \rangle &= 2 p_{0-} V_{\rm LF},\\
    \langle p_N q_1 \cdots q_N | p_N q_1 \cdots q_N \rangle &= 2 p_{N-} V_{\rm LF} \prod_{n=1}^N 2 \omega_{n} V,
\end{align*}
and 
\begin{equation*}
    |\mathcal{S}_N|^2 = (2 \pi)^3 \delta^3\left(   p_{N}-p_{0}+q_{1}^N \right)_{- \perp} V_{\rm LF} \frac{1}{2 \omega_0^2} |\mathcal{M}_N|^2.
\end{equation*}
Substituting these expressions into Eq. (\ref{eq:dW_N}), yields:
\begin{equation}
\begin{split}
    dW &= \frac{1}{2 (p_0 \cdot k) 2(p_N \cdot k)} |\mathcal{M}_N|^2 \frac{d^3 p_N}{(2 \pi)^3} \prod_{n=1}^N  \frac{d^3 \mathbf{q}_n}{(2 \pi)^3 2 \omega_{n} } \\
    & \times (2 \pi)^3 \delta^3\left(   p_{N}-p_{0}+q_{1 }^N \right)_{- \perp},
\end{split}
\end{equation}
 and after integrating out $p_N$, we have
\begin{equation}
    dW = \frac{1}{2 (p_0 \cdot k) 2(p_N \cdot k)} |\mathcal{M}_N|^2  \prod_{n=1}^N  \frac{d^3 \mathbf{q}_n}{(2 \pi)^3 2 \omega_{n} }.
\end{equation}
By rewriting $d^3 \mathbf{q}_n/ \omega_{n}  = \omega_n d \omega_n d \Omega_n$, we obtain the differential $N$-photon emission probability:
\begin{equation}\label{eq:dWN_FPW}
     \partial_{\omega, \Omega}^N W= \frac{1}{2 (p_0 \cdot k) 2(p_N \cdot k)} |\mathcal{M}_N|^2  \prod_{n=1}^N \frac{\omega_{n}}{2 (2 \pi)^3}.
\end{equation}
Here in analog with the differential two-photon emission probability, the differential $N$-photon emission probability can conveniently be normalized by the laser frequency: $\hat{\partial}_{\omega, \Omega}^N W = \omega_0^N \partial_{\omega, \Omega}^N W$.\\

\section{$N$-photon emission amplitude}\label{N_Photon_Emission_Amplitude}
Here we calculate the $N$-photon emission scattering amplitude. In this work we calculate tree-level Compton scattering amplitudes, therefore we only need to take into account the $N$\textsuperscript{th}-order of the S-matrix. The scattering matrix elements are given by:
\begin{equation*}
    \mathcal{S}_N  = \frac{1}{N!}\bra{p_N q_1 \cdots q_N} T\left\{\left( -ie\int \,d^4 x  \overline{\psi} \slashed{A} \psi \right)^N\right\} \ket{p_0}.
\end{equation*}
For the Wick contractions there are $N!$ identical contractions of electron fields, so this cancels out the $N!$ fraction. 
Noting that the Green's function of Eq. (\ref{eq:pulsedqqG}) has a propagator part of $ \slashed{p}_n -\frac{p_n^2-m^2}{2 k\cdot p_n}\slashed{k} \equiv \slashed{P}_n$ independent on $p_+$, we can compute all the integrals except for the $x_-$ integration, and by defining: $M_n^{\pi(1)}(\phi_n) = \slashed{\epsilon}_n^*-\frac{(\epsilon_n^* \cdot k) e^2 A_B^2(\phi_n)}{2(P_{n-1}^{\pi(1)}\cdot k)(P_n^{\pi(1)}\cdot k)}\slashed{k}+e \frac{\slashed{A}_B(\phi_n) \slashed{k} \slashed{\epsilon}_n^*  }{2 (P_n^{\pi(1)}\cdot k)} + e \frac{\slashed{\epsilon}_n^* \slashed{k}  \slashed{A}_B(\phi_n)}{2 (P_{n-1}^{\pi(1)}\cdot k)}$, $g_{n}^{\pi(j)}(\phi_n) = \kappa_{n}^{\pi(j)}\phi_n+ \int_{-\infty}^{\phi_n}\,d \phi' \left[ 2 e\Re\{\zeta_{n}^{\pi(j)} \mathcal{A}\} -e^2 \upsilon_{n}^{\pi(j)} A_B^2/2\right]$, $\zeta_n^{\pi(1)} = \frac{\epsilon\cdot (p_0-q_1^n)}{k\cdot (p_0-q_1^n)}-\frac{\epsilon\cdot (p_0-q_1^{n-1})}{k\cdot (p_0-q_1^{n-1})}$, $\upsilon_n^{\pi(1)} = \frac{1}{k\cdot (p_0-q_1^{n})}-\frac{1}{k\cdot (p_0-q_1^{n-1})}$ and $\kappa_n^{\pi(1)} = \frac{p_0\cdot q_1^{n}-(q_1^{n})^2/2}{k\cdot (p_0-q_1^{n})}-\frac{p_0\cdot q_1^{n-1}-(q_1^{n-1})^2/2}{k\cdot (p_0-q_1^{n-1})}$, we have
\begin{widetext}
\begin{equation}\label{eq:MN}
\begin{split}
         \mathcal{S}_N^{\pi(j)} &= (-ie)^N \int \,d (\phi_N,\phi_{N-1}\cdots,\phi_1) \exp\left[i g_N^{\pi(j)}(\phi_N)\right]  \overline{u}_{\sigma_N}(p_N) M_N^{\pi(j)}(\phi_N) \prod_{n=N-1}^1 \left(  \exp\left[ i g_n^{\pi(j)}(\phi_n) \right] \right. \\& \times \left.   \left[\left(\slashed{P}_n^{\pi(j)}+m \right) \left(\frac{\Theta(P_n^{\pi(j)}\cdot k)}{2 k\cdot P_n^{\pi(j)}}\Theta(\phi_{n+1}-\phi_n)+\frac{\Theta(-P_n^{\pi(j)}\cdot k)}{-2 k\cdot P_n^{\pi(j)}}\Theta(\phi_{n}-\phi_{n+1})\right)+ \frac{i \slashed{k}}{2 P_n^{\pi(j)}\cdot k}\delta(\phi_{n+1}-\phi_{n}) \right]  M_n^{\pi(j)}(\phi_n)\right)    \\&\times u_{\sigma_0}(p_0) (2\pi)^3 \delta\left(k\cdot\left[p_N-p_0+q_1^N \right] \right) \delta^2\left(   \mathbf{p}_{N\perp}-\mathbf{p}_{0\perp}+\mathbf{q}_{1 \perp}^N \right)\\&= i \mathcal{M}_N^{\pi(j)} (2\pi)^3 \delta\left(k\cdot\left[p_N-p_0+q_1^N \right] \right) \delta^2\left(   \mathbf{p}_{N\perp}-\mathbf{p}_{0\perp}+\mathbf{q}_{1 \perp}^N \right).
\end{split}
\end{equation}
\end{widetext}
The final momentum can be expressed as $p_N=p_0+\sum_{n=1}^N(\kappa_n k-q_n)$, or more general: $P_n^{\pi(1)} = P_{n-1}^{\pi(1)}+\kappa_n^{\pi(1)} k-q_n$. We choose $\pi(1)$ to be the permutation that $q_n$ corresponds to the $n$\textsuperscript{th} photon emission, one is free to order the other permutations. Note that there are $N!$ photon permutations for the $N$-photon emission scattering amplitude. The total scattering amplitude is the summation of all the photon permutations $\pi$, that is:
\begin{equation}\label{eq:MN2}
    i \mathcal{M}_N = \sum_{j=1}^{N!} i \mathcal{M}_N^{\pi(j)}.
\end{equation}

\section{Entanglement of a two qubit system}\label{Entanglement_of_a_Two_Qubit_System}

Below, we provide a more detailed description on how to apply the concurrence measure in a specific basis, and we agnostically refer to any 2-level quantum system as a 'qubit'. 

W.K. Wootters has shown that for an arbitrary 2-qubit mixed state with normalized density matrix $\hat{\rho}$, the concurrence is given by \cite{Concurrence}:
\begin{equation}
    C(\hat{\rho}) = \text{max}(0,\lambda_1-\lambda_2-\lambda_3-\lambda_4),
\end{equation}
such that $\lambda_i$ are the eigenvalues of the Hermitian matrix $F = \sqrt{\sqrt{\hat{\rho}}\tilde{\rho}\sqrt{\hat{\rho}}}$, in descending order. We define $\tilde{\rho} = (\sigma_2 \otimes \sigma_2) \hat{\rho}^* (\sigma_2 \otimes \sigma_2) $, and the fidelity is expressed as $\mathcal{F} = \left[\Tr\left(F\right)\right]^2$. Alternatively, one can take the square roots of the eigenvalues of $\hat{\rho} \tilde{\rho}$, as in Eq. (\ref{eq:concurrence1}). 

Since the photons are massless bosons, they have two polarization degrees of freedom and can therefore be represented by a two qubit system. If the final polarization is taken as a basis of the two qubit system and we take the basis
\begin{align*}
    \ket{0}_1 \otimes \ket{0}_2 &= \begin{pmatrix} 1 \\ 0 \\ 0 \\ 0 \end{pmatrix} & \ket{0}_1 \otimes \ket{1}_2 &= \begin{pmatrix} 0 \\ 1 \\ 0 \\ 0 \end{pmatrix},\\
     \ket{1}_1 \otimes \ket{0}_2 &= \begin{pmatrix} 0 \\ 0 \\ 1 \\ 0 \end{pmatrix} & \ket{1}_1 \otimes \ket{1}_2 &= \begin{pmatrix} 0 \\ 0 \\ 0 \\ 1 \end{pmatrix},
\end{align*}
where the subscripts represent the two photons and $\{\ket{0},\ket{1}\}$ is an orthogonal photon polarization basis. Then the density matrix is represented by \cite{E.Lotstedt_PRA}:
\begin{equation}\label{eq:rhofM}
\begin{split}
    \rho &= 
    \begin{pmatrix}
    \mathcal{M}_{00} \mathcal{M}_{00}^* & \mathcal{M}_{00} \mathcal{M}_{01}^* & \mathcal{M}_{00} \mathcal{M}_{10}^* & \mathcal{M}_{00} \mathcal{M}_{11}^*  \\ 
    \mathcal{M}_{01} \mathcal{M}_{00}^* & \mathcal{M}_{01} \mathcal{M}_{01}^* & \mathcal{M}_{01} \mathcal{M}_{10}^* & \mathcal{M}_{01} \mathcal{M}_{11}^*  \\
    \mathcal{M}_{10} \mathcal{M}_{00}^* & \mathcal{M}_{10} \mathcal{M}_{01}^* & \mathcal{M}_{10} \mathcal{M}_{10}^* & \mathcal{M}_{10} \mathcal{M}_{11}^*  \\
    \mathcal{M}_{11} \mathcal{M}_{00}^* & \mathcal{M}_{11} \mathcal{M}_{01}^* & \mathcal{M}_{11} \mathcal{M}_{10}^* & \mathcal{M}_{11} \mathcal{M}_{11}^* 
    \end{pmatrix} \\
    & \times \frac{ \omega_1 \omega_2  }{ 16 (2\pi)^6 (p_0 \cdot k) (p_2 \cdot k)} ,
\end{split}
\end{equation}
here the subscript is a chosen polarization basis where the first index is for photon $1$ and the second for photon $2$. Therefore, we have dropped the $2$ subscript (in $\mathcal{M}_2$), since it is clear from the photon basis. However, it is important to realize that the resulting concurrence is independent of the basis chosen. For the density matrix $\rho$ we average and sum the initial and final electron spin. We normalise the density matrix in the conventional manner:
\begin{equation}
  \hat{\rho} = \frac{1}{\Tr(\rho)} \rho.  
\end{equation}
In the definition of the density matrix of Eq. (\ref{eq:rhofM}), the trace of the density matrix is equal to the differential photon emission probability: $\Tr(\rho) = \partial_{\omega, \Omega}^2 W$.
The spin flip operator in this basis becomes:
\begin{equation}
    (\sigma_2 \otimes \sigma_2) = 
    \begin{pmatrix}
    0 & 0 & 0 & -1  \\ 
    0 & 0 & 1 & 0  \\
    0 & 1 & 0 & 0  \\
    -1 & 0 & 0 & 0 
    \end{pmatrix}.
\end{equation}
This shows that the entanglement is determined by the photon emission scattering amplitude. The polarization is also determined by the density matrix, hence all the information of the system (that is, regarding the emitted photons) is contained in $\rho$. 

\section{Polarization and Stokes parameters}\label{Ap.Polarization_and_Stokes_Parameters}

Here we give an approach to systematically calculate the Stokes parameters for a multiple photon system.  
Even though a basis independent and Lorentz invariant treatment of polarization is possible \cite{Gammie_2012}, this would make the calculation more convoluted than it needs to be. Therefore, for simplicity's sake, we will continue within the lab reference frame and choose a polarization basis. Here we use the laser propagation axis as the $\hat{z}$ axis and the $\hat{x}$ axis as the polarization axis of the laser, or $\hat{x}=\sqrt{2} \Re\{\epsilon\}$ for a linearly and circularly polarized laser, respectively. Then the photons' four-vectors for a $N$-photon system are:
\begin{align*}
    q_{j} &= \omega_{j} (1,\sin(\theta_{j}) \cos(\varphi_{j}),\sin(\theta_{j}) \sin(\varphi_{j}),\cos(\theta_{j})),\\
    \epsilon_{j 0} &= (0,\cos(\theta_{j}) \cos(\varphi_{j}),\cos(\theta_{j}) \sin(\varphi_{j}),-\sin(\theta_{j})),\\
    \epsilon_{j 1} &= (0,-\sin(\varphi_{j}),\cos(\varphi_{j}),0),
\end{align*}
such that $j\in \{1,2,\cdots,N\}$, $q_j\cdot \epsilon_{j0/1}=0$, $q_j^2=0$  and $\epsilon_{j0/1}^2=-1$, which directly shows that this is a linear polarization basis, this will be referred to as the $\{0,1\}$ basis. The circular basis can be defined as:
\begin{equation}
    \epsilon_{j \pm} =\frac{1}{\sqrt{2}}(\epsilon_{j 0}\pm i \epsilon_{j 1}),
\end{equation}
which, in turn, will be referred to as the $\{+,-\}$ basis.
As one might expect the polarization axis of the photons change with the emission angles. In experiment however, one would have to align the polarizers along the emission angle. A more convenient setup would be to place the polarizer in front of the Compton source. Therefore, we map these polarization vectors onto the polarization plane $x-y$, $\alpha \hat{x}+\beta \hat{y}\equiv (\alpha,\beta)$. First, we will start with mapping the $\{0,1\}$ basis. Since $\epsilon_{j1}$ is already fully in the polarization plane, we only need to map $\epsilon_{j0}$:
$$\boldsymbol{\epsilon}^{m}_{j0} = \frac{\boldsymbol{\epsilon}^{\perp}_{j0}}{\sqrt{\boldsymbol{\epsilon}^{\perp 2}_{j0}}}= (\cos(\varphi_{j}),\sin(\varphi_{j})), $$
where in the last step, we assume that we measure back-scattered photons $\theta_j < \pi/2$. We will denote photons polarized along the $\hat{x}$ direction as state $\ket{H}$ and photons polarized along the $\hat{y}$ direction as state $\ket{V}$. Transforming $\rho_N$ in the basis $\{H,V\}$ instead of $\{0,1\}$ as in Eq. (\ref{eq:rhofM}) would allow for a photon emission angle independent basis, where $\rho_N$ is a $N$-qubit density matrix (note that above we take $\rho_2\equiv \rho$). We normalize $\rho_N$ such that $\Tr\{\rho_N\}= \partial^N_{\omega,\Omega}W$ and define $\hat{\rho}_N = \rho_N/\Tr\{\rho_N\}$. The direct mapping between the two bases is the rotation matrix $R_j$:
\begin{equation}
    \begin{pmatrix}
        \ket{H}\\
        \ket{V}
    \end{pmatrix}_j
    = R_j 
    \begin{pmatrix}
        \ket{0}\\
        \ket{1}
    \end{pmatrix}_j,
\end{equation}
$$ R_j = 
\begin{pmatrix}
    \cos(\varphi_j) & -\sin(\varphi_j)\\
    \sin(\varphi_j) & \cos(\varphi_j)
\end{pmatrix}. 
$$
The basis transformation of the density matrix becomes:
\begin{equation}
    \rho_N^{\{H,V 
    \}} = \left(\bigotimes_{j=1}^N R_j\right)  \rho_N^{\{0,1 
    \}} \left(\bigotimes_{j=1}^N R_j\right)^\dag.
\end{equation}
In the case of double-Compton scattering we have two photons, $q_1$ and $q_2$. Therefore, the basis transformation of the $\rho_2$ is:
\begin{equation}
    \rho_2^{\{H,V 
    \}} = (R_1 \otimes R_2) \rho_2^{\{0,1 
    \}} (R_1 \otimes R_2)^\dag.
\end{equation}
To come back to the circular basis, to map the $\{+,-\}$ basis onto the polarization plane, we can use the mapped $\{0,1\}$ basis:
\begin{equation}
    \boldsymbol{\epsilon}_{j \pm}^m =\frac{1}{\sqrt{2}}(\boldsymbol{\epsilon}_{j 0}^m\pm i \boldsymbol{\epsilon}_{j 1}^\perp).
\end{equation}
Then we can transform this basis into the $\{L,R\}$ basis, defined by:
\begin{equation}
U_{j_1 j_2} = \langle \{H,V\}_{j_1} | \{L,R\}_{j_2}\rangle,
\end{equation}
where
\begin{equation*}
U=\frac{1}{\sqrt{2}}
     \begin{pmatrix}
       1 & 1 \\
       i & -i 
    \end{pmatrix}.
\end{equation*}
Such that $\ket{L} = (\ket{H}+i\ket{V})/\sqrt{2}$ and $\ket{R}= (\ket{H}-i\ket{V})/\sqrt{2}$ are the well known definitions of the left- and right-handed polarization photon states, respectively. The $\{L,R\}$ basis is obtained by the transformation:
\begin{equation}
    \rho_N^{\{L,R 
    \}} = \left(\bigotimes_{j=1}^N U^\dag R_j U\right)  \rho_N^{\{+,- 
    \}} \left(\bigotimes_{j=1}^N U^\dag R_j U\right)^\dag.
\end{equation}
To complete all the basis transformations, the transformation $\{H,V\}\rightarrow \{L,R\}$, is given by:
\begin{equation}
    \rho_N^{\{L,R 
    \}} = \left(\bigotimes_{j=1}^N U^{\dag}\right)  \rho_N^{\{H,V 
    \}} \left(\bigotimes_{j=1}^N U\right).
\end{equation}

\subsection{Stokes parameters}

The Stokes parameters can be constructed from the expectation values of the following operators:
\begin{align*}
    S_0 &= \ket{H}\bra{H}+\ket{V}\bra{V} &
    S_1 &= \ket{H}\bra{H}-\ket{V}\bra{V},\\
    S_2 &= \ket{V}\bra{H}+\ket{H}\bra{V} &
    S_3 &= i(\ket{V}\bra{H}-\ket{H}\bra{V}),
\end{align*}
or in the $\{H,V\}$ basis matrix representation:
\begin{align}
    S_0 &= \sigma_0 &
    S_1 &= \sigma_3, \\
    S_2 &= \sigma_1 & 
    S_3 &= \sigma_2.
\end{align}
In the circular basis $\{L,R\}$ this becomes:
\begin{equation}
    S_l^{\{L,R\}} = U^{\dag} S_l^{\{H,V\}} U = \sigma_l,
\end{equation}
which yields the Stokes parameters after taking the expectation value of the operators:
\begin{equation}
    s_l = \Tr\left\{\rho_1 S_l\right\}, 
\end{equation}
with $\rho_1$ being an one qubit density matrix, for example, the emitted photon from single-Compton scattering. Here the $s_0$ parameter is equal to the differential photon emission probability $s_0 = \partial^1_{\omega,\Omega} W$; the $s_1$ parameter is the degree of horizontal/vertical polarization; the $s_2$ parameter is the degree of diagonal polarization; the $s_3$ parameter is the degree of circular polarization. For a $N$-qubit system the generalized Stokes parameters can be defined by \cite{Stokes_1}:
\begin{equation}
    s_{l_1 l_2 \cdots l_N} = \Tr\left\{\rho_N \bigotimes_{i=1}^N S_{l_i} \right\}. 
\end{equation}
Here we refer to the generalized Stokes parameters as the $N$-qubit Stokes parameters.
For two photons this yields:
\begin{equation}
    s_{l_1 l_2 } = \Tr\left\{\rho_2  S_{l_1} \otimes S_{l_2} \right\}. 
\end{equation}
Where again $s_{00} = \Tr\{\rho_2\} = \partial_{\omega, \Omega}^2 W$. For convenience, the diagonal tensor products of $S_l$ in the matrix representation are given:
\begin{align*}
    \left(S_{0} \otimes S_{0}\right)^{\{H,V\}}=\left(S_{0} \otimes S_{0}\right)^{\{L,R\}} &= 
    \begin{pmatrix}
        1 & 0 & 0 & 0 \\
        0 & 1 & 0 & 0 \\
        0 & 0 & 1 & 0 \\
        0 & 0 & 0 & 1 
    \end{pmatrix}, \\
     \left(S_{1} \otimes S_{1}\right)^{\{H,V\}}=\left(S_{3} \otimes S_{3}\right)^{\{L,R\}} &= 
    \begin{pmatrix}
        1 & 0 & 0 & 0 \\
        0 & -1 & 0 & 0 \\
        0 & 0 & -1 & 0 \\
        0 & 0 & 0 & 1 
    \end{pmatrix}, \\
     \left(S_{2} \otimes S_{2}\right)^{\{H,V\}}=\left(S_{1} \otimes S_{1}\right)^{\{L,R\}} &= 
    \begin{pmatrix}
    0 & 0 & 0 & 1  \\ 
    0 & 0 & 1 & 0  \\
    0 & 1 & 0 & 0  \\
    1 & 0 & 0 & 0 
    \end{pmatrix}, \\ 
    \left(S_{3} \otimes S_{3}\right)^{\{H,V\}}=\left(S_{2} \otimes S_{2}\right)^{\{L,R\}} &= 
    \begin{pmatrix}
    0 & 0 & 0 & -1  \\ 
    0 & 0 & 1 & 0  \\
    0 & 1 & 0 & 0  \\
    -1 & 0 & 0 & 0 
    \end{pmatrix}.
\end{align*}
We normalise the Stokes parameters so that they have a value between $[-1,1]$. This is done by dividing the Stokes parameters by the differential photon emission probability, yielding:
\begin{equation}
    \hat{s}_{l_1 l_2 \cdots l_N} = \frac{s_{l_1 l_2 \cdots l_N}}{s_{00 \cdots 0}} = \Tr\left\{\hat{\rho}_N \bigotimes_{i=1}^N S_{l_i} \right\}.
\end{equation} 
In the one photon/qubit case the inequality $1 \geq \hat{s}_1^2+\hat{s}_2^2+\hat{s}_3^2$ holds, which is an equality for a fully polarized source. The parameter $p = \sqrt{\hat{s}_1^2+\hat{s}_2^2+\hat{s}_3^2}$ is a measure for the degree of polarization of a source, $p=1$ is a fully polarized source and $p=0$ is fully unpolarized. This inequality for generalizes to  (where Ref. \cite{Stokes_2} applies $N=2$):
\begin{equation*}
\begin{cases}
   1 \geq \sum_{l_1,l_2,\cdots l_N=1}^3 \hat{s}_{l_1 l_2 \cdots l_N}^2 = p^2 & \text{separable},\\
    1 \geq \frac{1}{2}\left(-1+\sum_{l_1,l_2,\cdots l_N=1}^3 \hat{s}_{l_1 l_2 \cdots l_N}^2\right)=P^2 & \text{2-entangled},
\end{cases}
\end{equation*}
where the second inequality is valid for a system where two qubits are entangled and the others are separable.
For separable states it generalizes trivially. However, entangled states can have a higher degree of polarization, due to their entangled nature, for example, $\ket{\psi} = (\ket{HV}-\ket{VH})/\sqrt{2}= i (\ket{LR}-\ket{RL})/\sqrt{2} $ has $\hat{s}_{11}=\hat{s}_{22}=\hat{s}_{33}=-1$, hence for $\ket{\psi}$: $P=C=1$. $P$ is therefore a measure of the entanglement of the system, but it is not necessary equal to the concurrence. If a higher degree of entanglement is present in the system (i.e., more than two qubits are entangled), then more complicated measures of entanglement have to be used, e.g. the $N$-tangle \cite{Three_tangle_1,Three_tangle_2,n_tangle,n_tangle_odd,n_tangle_2,three_tangle_3,n_tangle_3}.  

\begin{figure}[t]
    \centering
    \includegraphics[scale=0.7]{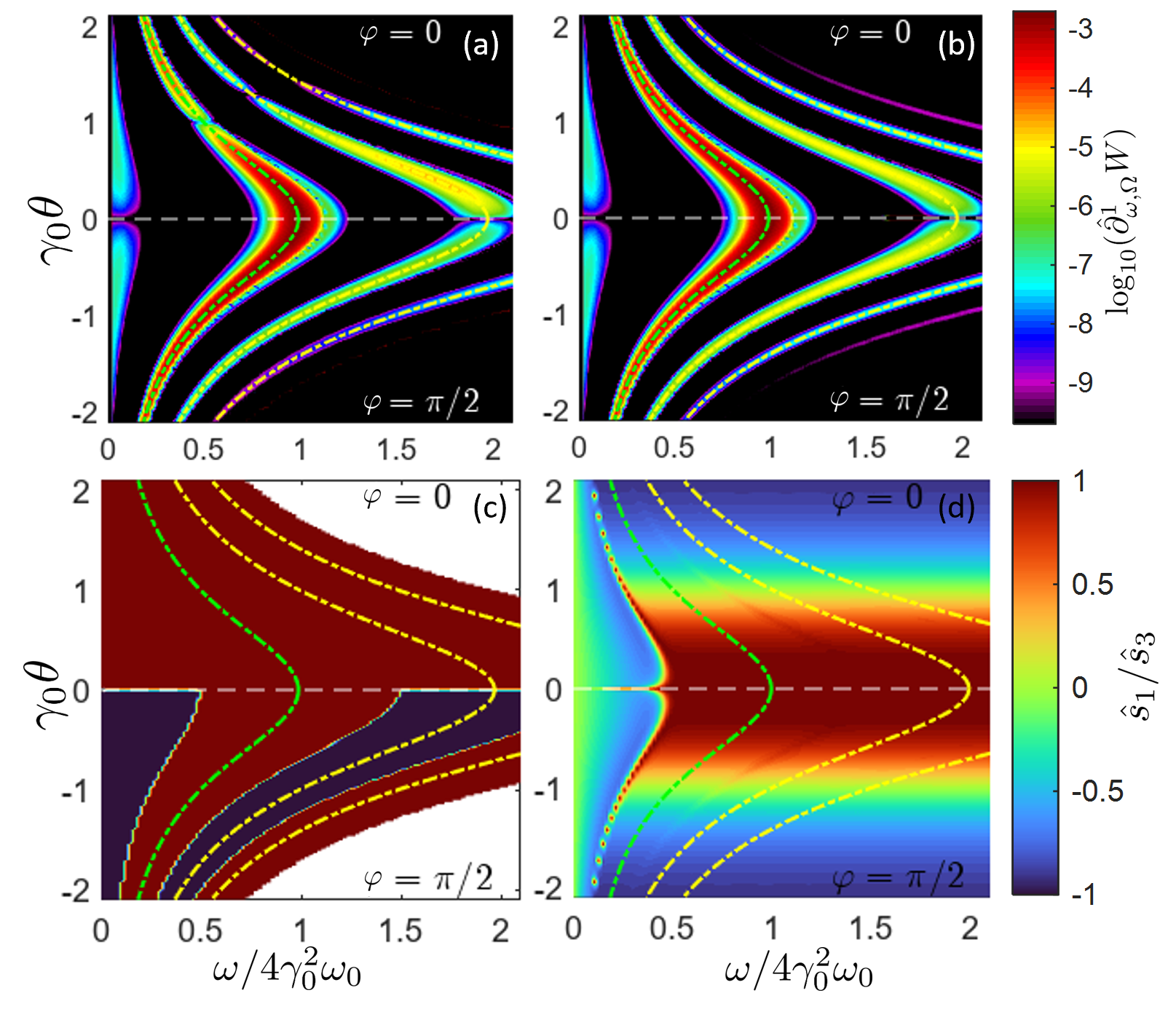}
    \captionsetup{justification=Justified}
    \caption{Differential photon emission and polarization for head-on scattering with the parameters: $a_0=0.1$, $\omega_0/m = 10^{-5}$, $\Delta \phi = 30 \pi$ and $\gamma_0 \approx 70.7$ ($|\boldsymbol{\beta}_0| = 0.9999$). Here, \textbf{(a)} depicts the normalized angular spectral photon emission probability $\log_{10}(\hat{\partial}_{\omega, \Omega}^1 W)$ for a linear polarized laser source $\ket{H}$. \textbf{(b)} depicts the normalized angular spectral photon emission probability $\log_{10}(\hat{\partial}_{\omega, \Omega}^1 W)$ for a right-handed circularly polarized laser source $\ket{R}$. \textbf{(c)} depicts the $\hat{s}_1$ Stokes parameter for the linearly polarized laser source ($\ket{H}$) and \textbf{(d)} depicts the $\hat{s}_3$ Stokes parameter for a right-handed circularly polarized laser source ($\ket{R}$). The green dashed line is the resonance condition of the fundamental and the yellow dashed lines of the higher harmonics. }
    \label{fig:1PhE_pol1}
\end{figure}

\section{Polarization of (single-)Compton scattering}\label{Polarization_of_Compton_Scattering}
We calculate the polarization of single-Compton scattering numerically by solving the trace of Eqs. (\ref{eq:MN}) and (\ref{eq:MN2}) for $N=1$, where we average and sum the initial and final electron spin, respectively. For the numerical calculations we consider head-on scattering [$\boldsymbol{\beta}_0= (0,0,\beta_0)$ and $\mathbf{\hat{n}}_0=(0,0,-1)$], where we use a linearly polarized laser pulse in the $\hat{x}$ direction [$\epsilon=(0,1,0,0)$] and a right-handed circularly polarized laser [$\epsilon=(0,1,-i,0)/\sqrt{2}$]. For the pulsing we use a cosine squared pulse of the form: $\mathcal{A}(\phi) = A_0 \cos^2(\pi \phi/\Delta\phi) (\Theta(\phi+\Delta\phi/2)-\Theta(\phi-\Delta\phi/2)) \exp[-i\phi]$.

\begin{figure}[t]
    \centering
    \includegraphics[scale=0.9]{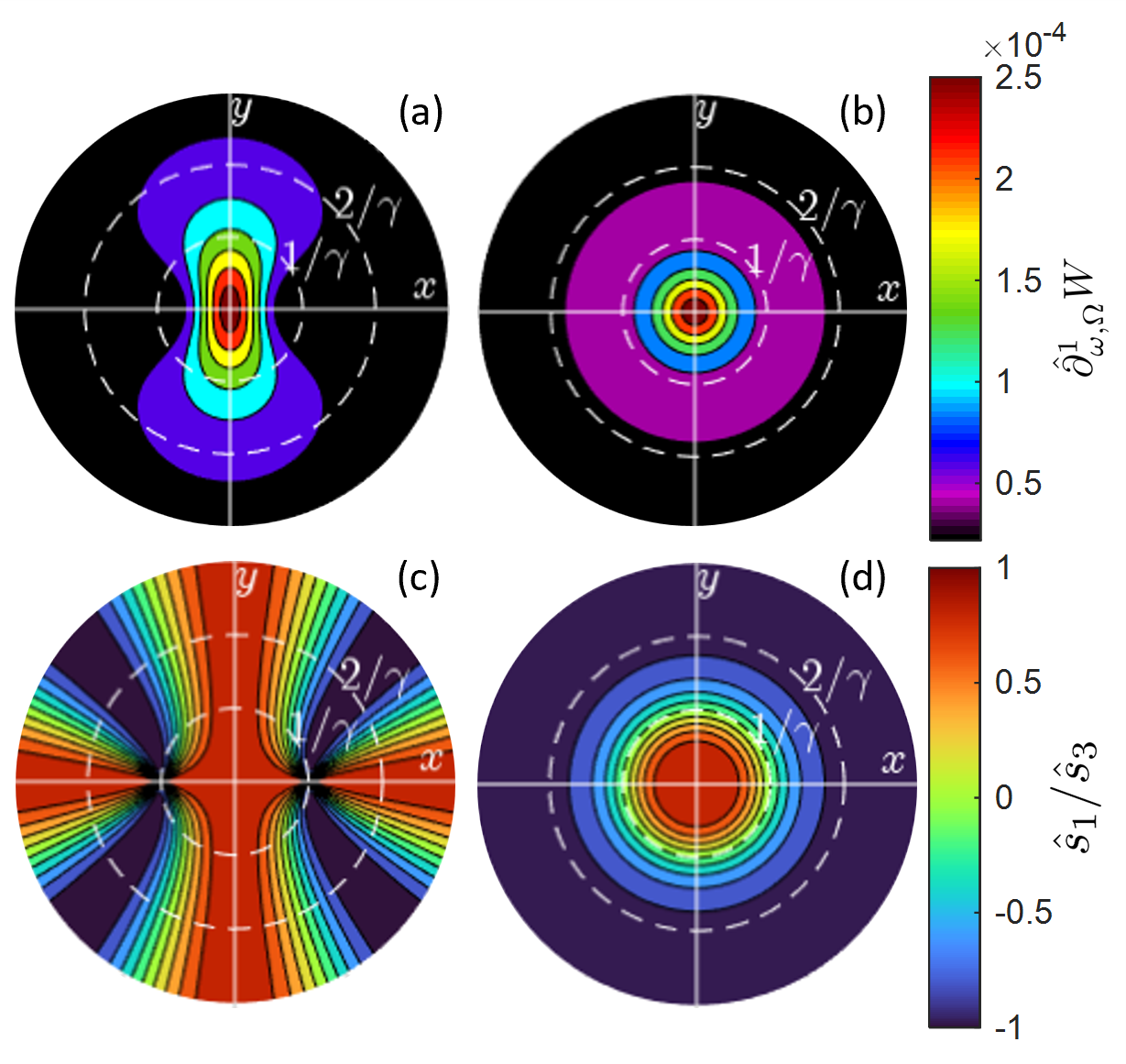}
    \captionsetup{justification=Justified}
    \caption{Polar Compton scattering plots for head-on scattering with parameters, $\gamma_0 \approx 70.7$ ($|\boldsymbol{\beta}_0| = 0.9999$), $\omega_0/m=10^{-5}$, $a_0=0.1$ and $\Delta \phi = 10 \pi$. \textbf{(a)} and \textbf{(b)} are the normalized differential photon emission probabilities $\hat{\partial}_{\omega, \Omega}^1 W$ for a linearly ($\ket{H}$) and circularly ($\ket{R}$) polarized laser source, respectively. \textbf{(c)} and \textbf{(d)} are the $\hat{s}_1$ and $\hat{s}_3$ Stokes parameters for a linearly and circularly polarized laser source, respectively. All plots are made of the fundamental mode, so the green dashed line of Fig. \ref{fig:1PhE_pol1}.}
    \label{fig:1PhE_pol2}
\end{figure}

Figures \ref{fig:1PhE_pol1} and \ref{fig:1PhE_pol2} show the polarization and photon emission probability of Compton scattering for $a_0=0.1$. Note that the Bremsstrahlung contribution (i.e. sub-harmonic component) is always linearly polarized, since this is due to the envelope of the laser pulse causing a longitudinal oscillation of the electron, by the ponderomotive interaction. For $\varphi=\pi/2$ of the linearly polarized laser source, the polarization interchanges for the $\hat{x}$ axis to the $\hat{y}$ axis and vice versa, this can be explained by the fact that the higher harmonics interchange from a transversal contribution to a longitudinal contribution, of the electron motion. The emission for a circularly polarized laser pulse is, on-axis, fully circularly polarized and becomes linearly polarized at $\theta = 1/\gamma_0$.

\section{Double Compton scattering amplitude analysis}\label{Double_Compton_Scattering_Amplitude_Analysis}
From Eq. (\ref{eq:MN}) the double-Compton scattering amplitude is obtained by substituting $N=2$, i.e. Eq. (\ref{eq:M2main}). We note that we have the following relations between the different electron momenta:
\begin{align*}
    P_1^{\pi(1)} &= p_0 -q_1 +\kappa_1^{\pi(1)} k = p_2 +q_2 -\kappa_2^{\pi(1)} k,\\
    P_1^{\pi(2)} &= p_0-q_2 + \kappa_1^{\pi(2)} k = p_2 +q_1-\kappa_2^{\pi(2)} k,\\
    p_2 &= p_0-q_1-q_2 + \tilde{\kappa} k ,
\end{align*}
where $\tilde{\kappa} = \kappa_1^{\pi(1)}+\kappa_2^{\pi(1)}= \kappa_1^{\pi(2)}+\kappa_2^{\pi(2)} $, in analog we define $\tilde{\zeta} = \zeta_1^{\pi(1)}+\zeta_2^{\pi(1)}$, $\tilde{\upsilon} = \upsilon_1^{\pi(1)}+\upsilon_2^{\pi(1)}$ and $\tilde{g}(\phi) = g_2^{\pi(1)}(\phi)+ g_1^{\pi(1)}(\phi)$. All the separate integrals can be written in two forms, where the first is: 
\begin{equation}
    \tilde{\mathcal{I}}_{j\pm} = \int\,d\phi \exp[\pm i \tilde{g}(\phi)]
    \begin{cases}
        1  & j= 0,\\
        \mathcal{\mathcal{A}}(\phi)/A_0  & j= 1,\\
        A_B^2(\phi)/A_0^2 & j=2,
    \end{cases}
\end{equation}
with the convention $\tilde{\mathcal{I}}_{j+} =\tilde{\mathcal{I}}_{j} $ and $A_0 = \text{max}\{ |\mathcal{A}(\phi)| \}_\phi$. The second integral form is:
\begin{equation*}
    \begin{split}
    \Upsilon_{j_2 j_1}^{\pi(j)}(l_2,l_1)  &= \int \,d \mathbf{\Phi}^{\pi(j)} A_{j_2}(\phi_2) A_{j_1}(\phi_1) \\ & \times \exp\left[i\left(l_2 \phi_2 +l_1 \phi_1+g_2^{\pi(j)}(\phi_2) +g_1^{\pi(j)}(\phi_1) \right)\right],
    \end{split}
\end{equation*}
here 
\begin{equation*}
\begin{split}
        \,d\mathbf{\Phi}^{\pi(j)}&=\,d(\phi_2,\phi_1)\left[\Theta(P_1^{\pi(j)}\cdot k)\Theta(\phi_{2}-\phi_1)\right. \\ & \left.-\Theta(-P_1^{\pi(j)}\cdot k)\Theta(\phi_{1}-\phi_{2})\right]
\end{split}
\end{equation*}
is the light-cone time ordered integration variable and:
\begin{equation*}
    A_{j}(\phi) = 
    \begin{cases} 1, &  j= 0,\\
    |\mathcal{A}|(\phi)/A_0, &  j= 1,\\
    A_B^2(\phi)/A_0^2, & j=2,
    \end{cases}
\end{equation*}
is the normalized laser field magnitude to the order $j$. It is straightforward to show that:
\begin{equation*}
 \tilde{\mathcal{I}}_0 = \frac{\tilde{\upsilon}}{2 \tilde{\kappa}} \tilde{\mathcal{I}}_2 -e A_0 \frac{ \tilde{\zeta}}{\tilde{\kappa}}\tilde{\mathcal{I}}_{1+}-e A_0 \frac{ \tilde{\zeta}^*}{\tilde{\kappa}}\tilde{\mathcal{I}}_{1-}^*.
\end{equation*}
From now on we will take into account the emitted photon basis $\epsilon_{1\alpha}$ for photon 1 and $\epsilon_{2\beta}$ for photon 2. With these integrals, the off-shell channel of the amplitude $i\mathcal{M}_{\alpha \beta}^{\rm Off}$ (we drop the $2$ subscript since it is clear by the photon emission basis, which amplitude is meant) can be written as:
\begin{widetext}
\begin{equation}
\begin{split}
    i \mathcal{M}^{\text{Off}}_{\alpha \beta}  &=  \frac{-e^2}{2(P_1^{\pi(1)} \cdot k)}\overline{u}_{\sigma_2}(p_2) \left[ U_{00}^{\pi(1)} \Upsilon_{00}^{\text{Off},\pi(1)}+U_{10}^{\pi(1)} \Upsilon_{10}^{\text{Off},\pi(1)}(-1,0)+U_{10}^{\pi(1)}(c) \Upsilon_{10}^{\text{Off},\pi(1)}(1,0)+U_{01}^{\pi(1)} \Upsilon_{01}^{\text{Off},\pi(1)}(0,-1)\right. \\& \left.+U_{01}^{\pi(1)}(c) \Upsilon_{01}^{\text{Off},\pi(1)}(0,1)  +U_{11}^{\pi(1)} \Upsilon_{11}^{\text{Off},\pi(1)}(-1,-1)+U_{11}^{\pi(1)}(c) \Upsilon_{11}^{\text{Off},\pi(1)}(1,1)+U_{11}^{'{\pi(1)}} \Upsilon_{11}^{\text{Off},\pi(1)}(1,-1)\right. \\& \left.+U_{11}^{'{\pi(1)}}(c) \Upsilon_{11}^{\text{Off},\pi(1)}(-1,1) +U_{20}^{\pi(1)} \Upsilon_{20}^{\text{Off},\pi(1)}+U_{02}^{\pi(1)} \Upsilon_{02}^{\text{Off},\pi(1)}  +U_{21}^{\pi(1)} \Upsilon_{21}^{\text{Off},\pi(1)}(0,-1) \right. \\& \left.+U_{21}^{\pi(1)}(c) \Upsilon_{21}^{\text{Off},\pi(1)}(0,1) +U_{12}^{\pi(1)} \Upsilon_{12}^{\text{Off},\pi(1)}(-1,0)+U_{12}^{\pi(1)}(c) \Upsilon_{12}^{\text{Off},\pi(1)}(1,0)  +U_{22}^{\pi(1)} \Upsilon_{22}^{\text{Off},\pi(1)}     \right] u_{\sigma_0}(p_0) \\& +\frac{-i e^2}{2(P_1^{\pi(1)} \cdot k)}\overline{u}_{\sigma_2}(p_2) \left[ D_0^{\pi(1)} \tilde{\mathcal{I}}_0 + D_{1+}^{\pi(1)} \tilde{\mathcal{I}}_{1+} + D_{1-}^{\pi(1)} \tilde{\mathcal{I}}_{1-}^*+ D_2^{\pi(1)} \tilde{\mathcal{I}}_{2} \right] u_{\sigma_0}(p_0)+\pi(2),
\end{split}
\end{equation}
when there is no argument for $l_1$ and $l_2$ we set them to zero by convention and $c = \epsilon \leftrightarrow \epsilon^*$. The $U$ and $D$ matrices are defined in Tabs. \ref{tab:Uij} and \ref{tab:Di}. 
Moreover, we define the off- and on-shell $\Upsilon$-integral separations, i.e.
\begin{align}
    \Upsilon_{j_2 j_1}^{\text{Off},\pi(j)}(l_2,l_1)  &= \int \,d (\phi_2,\phi_1) \frac{1}{2}\text{sgn}(\phi_2-\phi_1) A_{j_2}(\phi_2) A_{j_1}(\phi_1) \exp\left[i\left(l_2 \phi_2 +l_1 \phi_1+g_2^{\pi(j)}(\phi_2) +g_1^{\pi(j)}(\phi_1) \right)\right],\\
    \Upsilon_{j_2 j_1}^{\text{On},\pi(j)}(l_2,l_1)  &= \int \,d (\phi_2,\phi_1) \frac{1}{2} A_{j_2}(\phi_2) A_{j_1}(\phi_1) \exp\left[i\left(l_2 \phi_2 +l_1 \phi_1+g_2^{\pi(j)}(\phi_2) +g_1^{\pi(j)}(\phi_1) \right)\right],
\end{align}
such that $\Upsilon_{j_2 j_1}^{\pi(j)}(l_2,l_1)= \Upsilon_{j_2 j_1}^{\text{Off},\pi(j)}(l_2,l_1)+\Upsilon_{j_2 j_1}^{\text{On},\pi(j)}(l_2,l_1) $.
\begin{table}[H]
    \centering
    \caption{$U_{ij}$ Matrix Elements}
    \begin{tabular}{||M|M||}\hline
      U_{00}^{\pi(1)}   & \slashed{\epsilon}_{2 \beta}^* (\slashed{P}_1^{\pi(1)} +m) \slashed{\epsilon}_{1 \alpha}^* \\
      U_{10}^{\pi(1)} & e A_0 \left( \frac{\slashed{\epsilon} \slashed{k} \slashed{\epsilon}_{2 \beta}^*}{2(p_2 \cdot k)}+ \frac{\slashed{\epsilon}_{2 \beta}^* \slashed{k}\slashed{\epsilon}}{2(P_1^{\pi(1)} \cdot k)} \right) (\slashed{P}_1^{\pi(1)} +m) \slashed{\epsilon}_{1 \alpha}^* \\ U_{01}^{\pi(1)} & e A_0 \slashed{\epsilon}_{2 \beta}^* (\slashed{P}_1^{\pi(1)} +m) \left( \frac{\slashed{\epsilon} \slashed{k} \slashed{\epsilon}_{1 \alpha}^*}{2(P_1^{\pi(1)} \cdot k)}+ \frac{\slashed{\epsilon}_{1 \alpha}^* \slashed{k}\slashed{\epsilon}}{2(p_0 \cdot k)} \right) \\ U_{11}^{\pi(1)} & e^2 A_0^2 \left( \frac{\slashed{\epsilon} \slashed{k} \slashed{\epsilon}_{2 \beta}^*}{2(p_2 \cdot k)}+ \frac{\slashed{\epsilon}_{2 \beta}^* \slashed{k}\slashed{\epsilon}}{2(P_1^{\pi(1)} \cdot k)} \right) (\slashed{P}_1^{\pi(1)} +m) \left( \frac{\slashed{\epsilon} \slashed{k} \slashed{\epsilon}_{1 \alpha}^*}{2(P_1^{\pi(1)} \cdot k)}+ \frac{\slashed{\epsilon}_{1 \alpha}^* \slashed{k}\slashed{\epsilon}}{2(p_0 \cdot k)} \right) \\ U_{11}^{'\pi(1)} & e^2 A_0^2 \left( \frac{\slashed{\epsilon}^* \slashed{k} \slashed{\epsilon}_{2 \beta}^*}{2(p_2 \cdot k)}+ \frac{\slashed{\epsilon}_{2 \beta}^* \slashed{k}\slashed{\epsilon}^*}{2(P_1^{\pi(1)} \cdot k)} \right) (\slashed{P}_1^{\pi(1)} +m) \left( \frac{\slashed{\epsilon} \slashed{k} \slashed{\epsilon}_{1 \alpha}^*}{2(P_1^{\pi(1)} \cdot k)}+ \frac{\slashed{\epsilon}_{1 \alpha}^* \slashed{k}\slashed{\epsilon}}{2(p_0 \cdot k)} \right) \\ U_{20}^{\pi(1)} & -e^2 A_0^2 \frac{(\epsilon_{2 \beta}^* \cdot k)}{2 (p_2 \cdot k)(P_1^{\pi(1)} \cdot k)} \slashed{k}(\slashed{P}_1^{\pi(1)} +m)\slashed{\epsilon}_{1 \alpha}^* \\ U_{02}^{\pi(1)} & - e^2 A_0^2 \frac{(\epsilon_{1 \alpha}^*\cdot k)}{2(p_0 \cdot k)(P_1^{\pi(1)} \cdot k)}\slashed{\epsilon}_{2 \beta}^* (\slashed{P}_1^{\pi(1)} +m)\slashed{k} \\ U_{21}^{\pi(1)} & -e^3 A_0^3 \frac{(\epsilon_{2 \beta}^* \cdot k)}{2 (p_2 \cdot k)(P_1^{\pi(1)} \cdot k)} \slashed{k}(\slashed{P}_1^{\pi(1)} +m) \left( \frac{\slashed{\epsilon} \slashed{k} \slashed{\epsilon}_{1 \alpha}^*}{2(P_1^{\pi(1)} \cdot k)}+ \frac{\slashed{\epsilon}_{1 \alpha}^* \slashed{k}\slashed{\epsilon}}{2(p_0 \cdot k)} \right) \\ U_{12}^{\pi(1)} &  - e^3 A_0^3 \frac{(\epsilon_{1 \alpha}^*\cdot k)}{2(p_0 \cdot k)(P_1^{\pi(1)} \cdot k)} \left( \frac{\slashed{\epsilon} \slashed{k} \slashed{\epsilon}_{2 \beta}^*}{2(p_2 \cdot k)}+ \frac{\slashed{\epsilon}_{2 \beta}^* \slashed{k}\slashed{\epsilon}}{2(P_1^{\pi(1)} \cdot k)} \right) (\slashed{P}_1^{\pi(1)} +m)\slashed{k} \\ U_{22}^{\pi(1)} & e^4 A_0^4 \frac{(\epsilon_{1 \alpha}^* \cdot k)(\epsilon_{2 \beta}^*\cdot k)}{2 (p_0\cdot k)(p_2 \cdot k)(P_1^{\pi(1)} \cdot k)} \slashed{k} \\ \hline
    \end{tabular}
    \label{tab:Uij}
\end{table}

\begin{table}[H]
    \centering
    \caption{$D_{i}$ Matrix elements.}
    \begin{tabular}{||M|M||M|M||}\hline
      D_0^{\pi(1)} & \slashed{\epsilon}_{2 \beta}^* \slashed{k} \slashed{\epsilon}_{1 \alpha}^* &
      D_{1+}^{\pi(1)} & e A_0  \frac{\epsilon_{2 \beta}^*\cdot k}{p_2 \cdot k} \slashed{\epsilon} \slashed{k}\slashed{\epsilon}_{1 \alpha}^* +e A_0  \frac{\epsilon_{1 \alpha}^*\cdot k}{p_0 \cdot k} \slashed{\epsilon}_{2 \beta}^* \slashed{k}\slashed{\epsilon} \\
      D_{1-}^{\pi(1)} & e A_0  \frac{\epsilon_{2 \beta}^*\cdot k}{p_2 \cdot k} \slashed{\epsilon^*} \slashed{k}\slashed{\epsilon}_{1 \alpha}^* +e A_0  \frac{\epsilon_{1 \alpha}^*\cdot k}{p_0 \cdot k} \slashed{\epsilon}_{2 \beta}^* \slashed{k}\slashed{\epsilon^*} &
      D_2^{\pi(1)} & - e^2 A_0^2 \frac{(\epsilon_{2\beta}^* \cdot k)(\epsilon_{1 \alpha}^* \cdot k)}{(p_0 \cdot k)(p_2 \cdot k)} \slashed{k} \\ \hline
    \end{tabular}
    \label{tab:Di}
\end{table}

The on-shell contribution can be expressed as:
\begin{equation}
\begin{split}
    i \mathcal{M}^{\text{On}}_{\alpha \beta}  &=  \frac{-e^2}{2|P_1^{\pi(1)} \cdot k|}\overline{u}_{\sigma_2}(p_2) \left[ U_{00}^{\pi(1)} \Upsilon_{00}^{\text{On},\pi(1)}+U_{10}^{\pi(1)} \Upsilon_{10}^{\text{On},\pi(1)}(-1,0)+U_{10}^{\pi(1)}(c) \Upsilon_{10}^{\text{On},\pi(1)}(1,0)+U_{01}^{\pi(1)} \Upsilon_{01}^{\text{On},\pi(1)}(0,-1)\right. \\& \left.+U_{01}^{\pi(1)}(c) \Upsilon_{01}^{\text{On},\pi(1)}(0,1)  +U_{11}^{\pi(1)} \Upsilon_{11}^{\text{On},\pi(1)}(-1,-1)+U_{11}^{\pi(1)}(c) \Upsilon_{11}^{\text{On},\pi(1)}(1,1)+U_{11}^{'{\pi(1)}} \Upsilon_{11}^{\text{On},\pi(1)}(1,-1)\right. \\& \left.+U_{11}^{'{\pi(1)}}(c) \Upsilon_{11}^{\text{On},\pi(1)}(-1,1) +U_{20}^{\pi(1)} \Upsilon_{20}^{\text{On},\pi(1)}+U_{02}^{\pi(1)} \Upsilon_{02}^{\text{On},\pi(1)}  +U_{21}^{\pi(1)} \Upsilon_{21}^{\text{On},\pi(1)}(0,-1) \right. \\& \left.+U_{21}^{\pi(1)}(c) \Upsilon_{21}^{\text{On},\pi(1)}(0,1) +U_{12}^{\pi(1)} \Upsilon_{12}^{\text{On},\pi(1)}(-1,0)+U_{12}^{\pi(1)}(c) \Upsilon_{12}^{\text{On},\pi(1)}(1,0)  +U_{22}^{\pi(1)} \Upsilon_{22}^{\text{On},\pi(1)}     \right] u_{\sigma_0}(p_0) \\& +\pi(2).
\end{split}
\end{equation}
Note that $i\mathcal{M}_{\alpha \beta}=  i \mathcal{M}^{\text{Off}}_{\alpha \beta}+i \mathcal{M}^{\text{On}}_{\alpha \beta}$. It has been shown that the on-shell contribution can be manipulated into the expression \cite{Seipt_thesis}: 
\begin{equation}
\begin{split}
    i\mathcal{M}^{\text{On}}_{\alpha \beta} &= \frac{1}{2| k\cdot P_1^{\pi(1)}|}\frac{1}{2}\sum_{\sigma_1} (-ie) \overline{u}_{\sigma_2}(p_2) \int \,d \phi_2 \exp\left[i g_2^{\pi(1)}(\phi_2)\right]   M_2^{\pi(1)}(\phi_2) u_{\sigma_1}(P_1) \\& \times (-ie) \overline{u}_{\sigma_1}(P_1) \int \,d \phi_1 \exp\left[i g_1^{\pi(1)}(\phi_1)\right]   M_1^{\pi(1)}(\phi_1)   u_{\sigma_0}(p_0) + \pi(2) \\ & \equiv \frac{1}{4| k\cdot P_1^{\pi(1)}|} \mathcal{M}_{f \beta}^{\pi(1)} \mathcal{M}_{i \alpha}^{\pi(1)} + \pi(2),
\end{split}
\end{equation}
\end{widetext}
which shows the on-shell contribution is nothing more that the multiplication of two single-Compton scattering amplitudes, with a summation over the intermediate electron spin. Furthermore, the $f,i$ subscripts are to distinguish between the final and initial one-photon emission matrix elements.
 
It is interesting to ascertain whether the on-shell contribution can be entangled. As a consequence of the decoupling of the on-shell contribution, the concurrence of on-shell two-photon emission can be approximated as (for $r \ll 1$):
\begin{equation}\label{eq:OnC}
     C \approx  \frac{|\mathcal{M}_{f1}^{\pi(2)}\mathcal{M}_{i0}^{\pi(1)}-\mathcal{M}_{f0}^{\pi(2)}\mathcal{M}_{i1}^{\pi(1)}|}{2\sqrt{2}|P_1^{\pi(1)} \cdot k| |\mathcal{M}_2^{\rm On}|}(\pi(1) \leftrightarrow \pi(2)).
\end{equation}
When the recoil is small ($r \ll 1$) the electron momentum does not change significantly after the first emission, and therefore the two photon emissions are independent from each other and $\mathcal{M}_{f1(0)}^{\pi(2)} \approx \mathcal{M}_{i1(0)}^{\pi(1)} $ (recall that the sum over the intermediate electron spin is carried out), which makes the concurrence of Eq. (\ref{eq:OnC}) vanish. Additionally, this indicates that for high recoil the on-shell photon emission may become entangled due to the higher degree of inter-dependence of the two photon emission events.
\vspace{0.5cm}
\subsection{Integral relations}

In similar vein to (single-)Compton scattering, the zeroth order integrals of double-Compton scattering are ill-behaved. However, the zeroth order integrals can be expressed in terms of the higher orders \cite{F.Mackenroth_Double_Compton_scattering,Daniel_Seipt}. Here we give a direct computation of the dependent integrals to solve this ill-behaved nature. During the integration manipulation we will make use of the integral relation (for $\kappa \neq 0$): 
\begin{equation}\label{eq:int1}
    \int_{\delta}^\infty \,d\phi e^{i \kappa \phi} = \frac{i}{\kappa}e^{i \kappa \delta}.
\end{equation} 
Consider the following integral:
\begin{widetext}
\begin{align*}
    \int\,d\phi_1 e^{i g_1(\phi_1)-i\delta_{j 1}\phi_1} A_j(\phi_1) \int_{\phi_1}^\infty \,d\phi_2  e^{i g_2(\phi_2)} \partial_{\phi_2} g_2(\phi_2) &= \int\,d\phi_1 e^{i g_1(\phi_1)-i\delta_{j 1}\phi_1} A_j(\phi_1) \int_{g_2(\phi_1)}^\infty \,d g_2 e^{i g_2},\\
    &= i \int\,d\phi_1 e^{i g_1(\phi_1)+i g_2(\phi_1)-i\delta_{j 1}\phi_1} A_j(\phi_1),\\
    &= i \int\,d\phi_1 e^{i \tilde{g}(\phi_1)-i\delta_{j 1}\phi_1} A_j(\phi_1) = i \tilde{\mathcal{I}}_{j},
\end{align*}
and in a similar manner:
\begin{align*}
    \int\,d\phi_1 e^{i g_1(\phi_1)-i\delta_{j 1}\phi_1} A_j(\phi_1) \int_{-\infty}^{\phi_1} \,d\phi_2  e^{i g_2(\phi_2)} \partial_{\phi_2} g_2(\phi_2) &= \int\,d\phi_1 e^{i g_1(\phi_1)-i\delta_{j 1}\phi_1} A_j(\phi_1) \int^{g_2(\phi_1)}_{-\infty} \,d g_2 e^{i g_2},\\
    &= -i \int\,d\phi_1 e^{i g_1(\phi_1)+i g_2(\phi_1)-i\delta_{j 1}\phi_1} A_j(\phi_1),\\
    &= -i \int\,d\phi_1 e^{i \tilde{g}(\phi_1)-i\delta_{j 1}\phi_1} A_j(\phi_1) = -i \tilde{\mathcal{I}}_{j},
\end{align*}
therefore this concludes that we can write:
\begin{equation*}
    \int\,d\mathbf{\Phi}^{\pi(1)/\pi(2)} e^{i g_1^{\pi(1)/\pi(2)}(\phi_1)+ig_2^{\pi(1)/\pi(2)}(\phi_2)-i\delta_{j 1}\phi_1} A_j(\phi_1)   \partial_{\phi_2} g_2^{\pi(1)/\pi(2)}(\phi_2) = i \tilde{\mathcal{I}}_{j}\left[\Theta(P_{1}^{\pi(1)/\pi(2)}\cdot k)+\Theta(-P_1^{\pi(1)/\pi(2)}\cdot k)\right] = i \tilde{\mathcal{I}}_{j}.
\end{equation*}
From now on we will exclude the $\pi(1)/\pi(2)$ superscripts to keep the notation clean. Note that the same integral can evaluate to the following:
\begin{align*}
    i \tilde{\mathcal{I}}_{j} &= \int\,d\mathbf{\Phi} e^{i g_1(\phi_1)+ig_2(\phi_2)-i\delta_{j 1}\phi_1} A_j(\phi_1)  \partial_{\phi_2} g_2(\phi_2), \\
    &= \int\,d\mathbf{\Phi} e^{i g_1(\phi_1)+ig_2(\phi_2)-i\delta_{j 1}\phi_1} A_j(\phi_1) \left[\kappa_{2}+2 e\Re\{\zeta_{2} \mathcal{A}(\phi_2)\} -\frac{1}{2}e^2 \upsilon_{2} A_B^2(\phi_2) \right],\\
    &= \kappa_2 \Upsilon_{0 j}(0,-\delta_{j 1})-\frac{m a_0 \zeta_2}{\sqrt{2}}\Upsilon_{1 j}(-1,-\delta_{j 1})-\frac{m a_0 \zeta_2^*}{\sqrt{2}}\Upsilon_{1 j}(1,-\delta_{j 1})-\frac{m^2 a_0^2 \upsilon_2}{4} \Upsilon_{2 j}(0,-\delta_{j 1}),
\end{align*}
which after rearrangement reads:
\begin{equation}\label{eq:upsint1}
    \Upsilon_{0 j}(0,-\delta_{j 1}) = \frac{i}{\kappa_2} \tilde{\mathcal{I}}_{j}+\frac{m a_0 \zeta_2}{\sqrt{2}\kappa_2}\Upsilon_{1 j}(-1,-\delta_{j 1})+\frac{m a_0 \zeta_2^*}{\sqrt{2}\kappa_2}\Upsilon_{1 j}(1,-\delta_{j 1})+\frac{m^2 a_0^2 \upsilon_2}{4 \kappa_2} \Upsilon_{2 j}(0,-\delta_{j 1}).
\end{equation}
A similar integral of the form:
$$\int\,d\mathbf{\Phi} e^{i g_1(\phi_1)+i g_2(\phi_2)+i\phi_1} A_1(\phi_1)  \partial_{\phi_2} g_2(\phi_2),$$
can be evaluated in an identical manner yielding:
\begin{equation}\label{eq:upsint2}
    \Upsilon_{0 1}(0,1) = \frac{i}{\kappa_2} \tilde{\mathcal{I}}_{1-}^*+\frac{m a_0 \zeta_2}{\sqrt{2}\kappa_2}\Upsilon_{1 1}(-1,1)+\frac{m a_0 \zeta_2^*}{\sqrt{2}\kappa_2}\Upsilon_{1 1}(1,1)+\frac{m^2 a_0^2 \upsilon_2}{4 \kappa_2} \Upsilon_{2 1}(0,1).
\end{equation}
The other integral expressions can be obtained by evaluating:
$$\int\,d\mathbf{\Phi} e^{i g_1(\phi_1)+ig_2(\phi_2) \pm i\delta_{j 1}\phi_2} A_j(\phi_2)  \partial_{\phi_1} g_1(\phi_1). $$
Doing so yields:
\begin{equation}\label{eq:upsint3}
    \Upsilon_{j 0}(-\delta_{j 1},0) = -\frac{i}{\kappa_1} \tilde{\mathcal{I}}_{j}+\frac{m a_0 \zeta_1}{\sqrt{2}\kappa_1}\Upsilon_{j 1}(-\delta_{j 1},-1)+\frac{m a_0 \zeta_1^*}{\sqrt{2}\kappa_1}\Upsilon_{j 1}(-\delta_{j 1},1)+\frac{m^2 a_0^2 \upsilon_1}{4 \kappa_1} \Upsilon_{j 2}(-\delta_{j 1},0),
\end{equation}
\begin{equation}\label{eq:upsint4}
    \Upsilon_{1 0}(1,0) = -\frac{i}{\kappa_1} \tilde{\mathcal{I}}_{1-}^*+\frac{m a_0 \zeta_1}{\sqrt{2}\kappa_1}\Upsilon_{1 1}(1,-1)+\frac{m a_0 \zeta_1^*}{\sqrt{2}\kappa_1}\Upsilon_{1 1}(1,1)+\frac{m^2 a_0^2 \upsilon_1}{4 \kappa_1} \Upsilon_{1 2}(1,0).
\end{equation}
Which shows that all zeroth order integral expressions can be expressed in terms of the higher orders. We note that the same integral relations can be obtained by making use of the Ward identity (see e.g. \cite{Seipt_thesis}). 
\newpage
\end{widetext}

\subsection{On-shell and off-shell results}

\begin{figure}[H]
    \centering
    \includegraphics[scale=0.65]{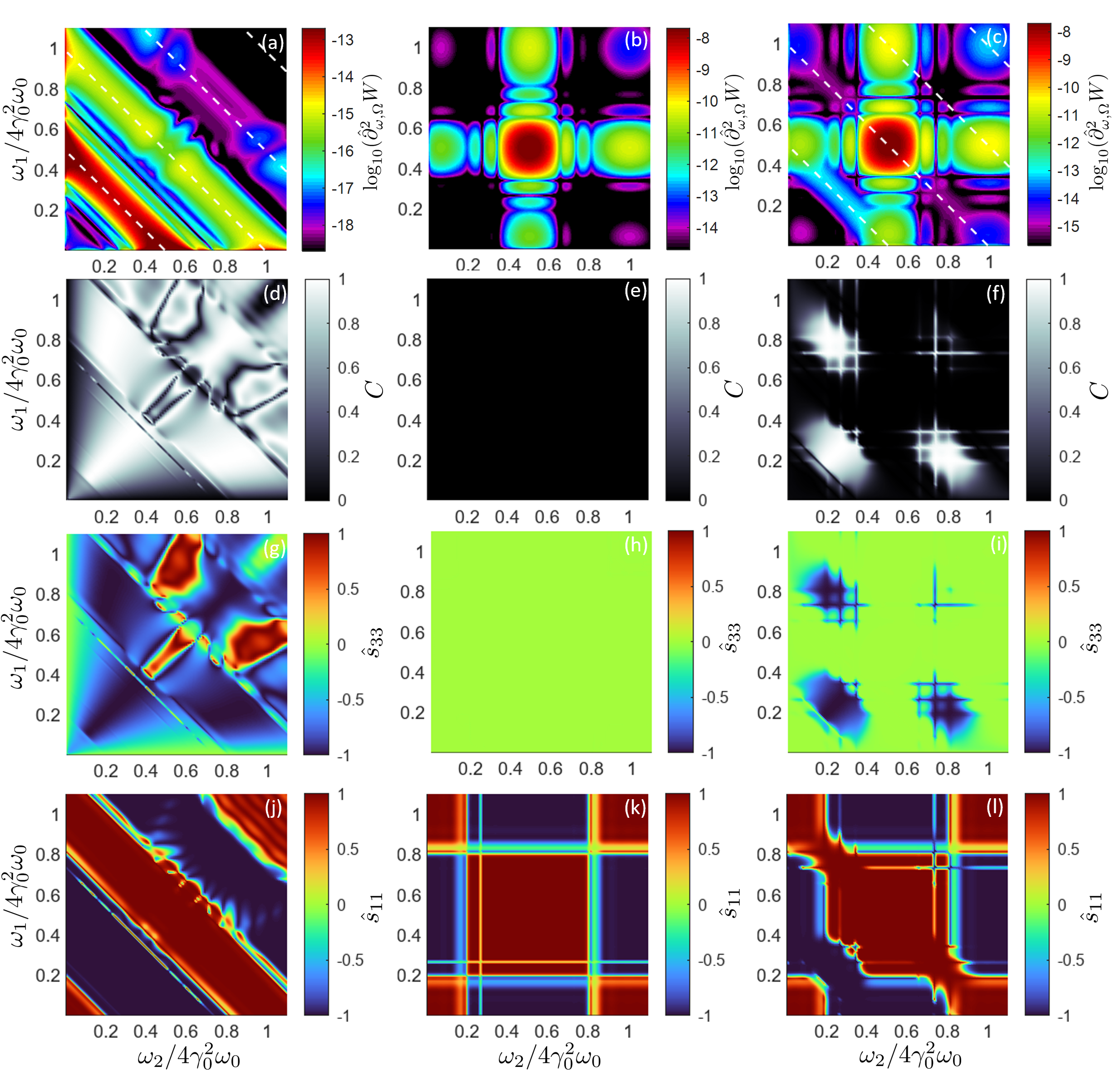}
    \caption{The spectral distribution of double-Compton scattering for the off-shell \textbf{(a,d,g,j)}, on-shell \textbf{(b,e,h,k)} and total contributions \textbf{(c,f,i,l)}. For an electron scattered head-on with a linearly polarized laser with $a_0=0.1$, $\gamma_0 \approx 70.7$ ($|\boldsymbol{\beta}_0| = 0.9999$), $\omega_0/m=10^{-5}$, $\Delta \phi = 40$, $\theta_1=\theta_2=1/\gamma_0$, $\varphi_1 = \pi/2$ and $\varphi_2 = 3\pi/2$. Here \textbf{(a-c)} depicts the normalized differential photon emission probability $\log_{10}(\hat{\partial}_{\omega, \Omega}^2 W)$, where the white dashed lines are the $s=\{1,2,3,4\}$ photon emission resonance frequencies, \textbf{(d-f)} the concurrence, \textbf{(g-i)} the generalized degree of circular polarization $\hat{s}_{33}$, \textbf{(j-l)} the generalized degree of linear polarization $\hat{s}_{11}$. }
    \label{fig:a0_01_FPW}
\end{figure}

The photon emission as discussed in the article is focused on the off-shell contribution of a linearly polarized laser source [$\epsilon = (0,1,0,0)$] for the orientation $\{\varphi_1=\pi/2,\varphi_2=3 \pi/2\}$ in a head-on scattering geometry[($\boldsymbol{\beta}_0= (0,0,\beta_0)$ and $\mathbf{\hat{n}}_0=(0,0,-1)$]. Here in addition, we present the on-shell and total contributions of the differential photon emission for a laser field with a longitudinal pulsing of $\mathcal{A}(\phi) = A_0 \cos^2(\pi \phi/\Delta\phi) (\Theta(\phi+\Delta\phi/2)-\Theta(\phi-\Delta\phi/2)) \exp[-i\phi]$.
Figure \ref{fig:a0_01_FPW} depicts the complete result, of a linearly polarized laser with a classical nonlinearity parameter of $a_0=0.1$, at the angles $\varphi_1=\pi/2$ and $\varphi_2=3 \pi/2$. As illustrated the total differential photon emission probability is approximately the differential photon emission probability of the on-shell and off-shell contributions combined. This is due to the fact that the on-shell and off-shell scattering amplitudes (in the regimes of interest) do not (significantly) interfere with each other $|\mathcal{M}^{\text{On}}+\mathcal{M}^{\text{Off}}|^2 \approx |\mathcal{M}^{\text{On}}|^2 +|\mathcal{M}^{\text{Off}}|^2$ \cite{Seipt_thesis}. For this reason the total concurrence and Stokes parameters are the differential photon emission probability weighted average of the off- and on-shell contributions. As depicted in Fig. \ref{fig:a0_01_FPW} the on-shell contribution is fully unentangled and has the same polarization as for two consecutive single-Compton/Thomson scattering events. For a non-destructive filtering, the frequencies should be filtered out where the on-shell contribution does not radiate, visible in Fig. \ref{fig:a0_01_FPW}\textcolor{blue}{.f}. 

\begin{figure}[H]
    \centering
    \includegraphics[scale=0.7]{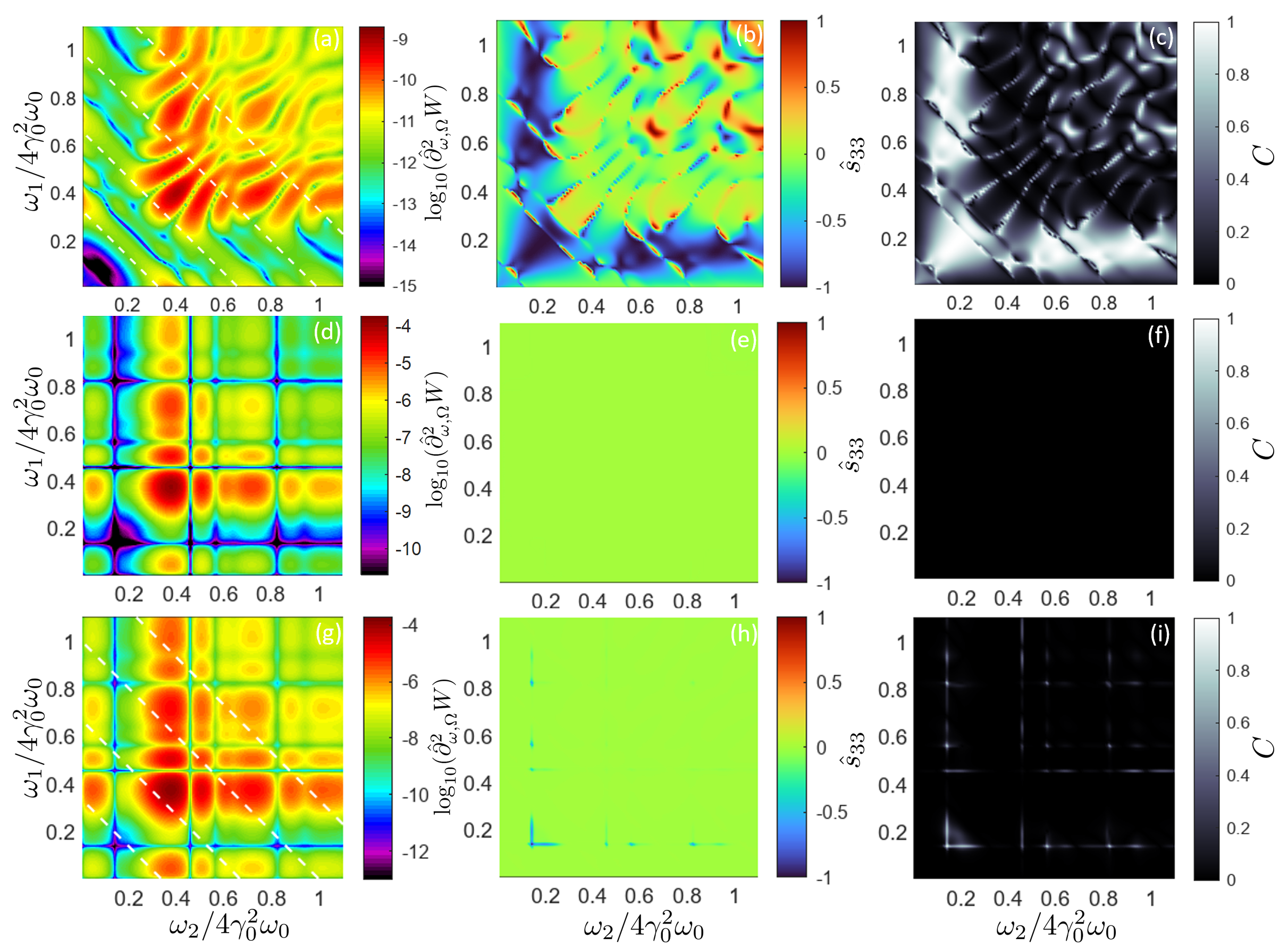}
    \caption{The spectral distribution of double-Compton scattering for the off-shell \textbf{(a-c)}, on-shell \textbf{(d-f)} and the total contributions \textbf{(g-i)}. For an electron scattered head-on with a linearly polarized laser with $a_0=1$, $\gamma_0 \approx 70.7$ ($|\boldsymbol{\beta}_0| = 0.9999$), $\omega_0/m=10^{-5}$, $\Delta \phi = 40$, $\theta_1=\theta_2=1/\gamma_0$, $\varphi_1 = \pi/2$ and $\varphi_2 = 3\pi/2$. Here \textbf{(a,d,g)} depicts the differential photon emission probability $\log_{10}(\hat{\partial}_{\omega, \Omega}^2 W)$, where the white dashed lines are the $s=\{1,2,3,4\}$ photon emission resonance frequencies, \textbf{(b,e,h)}  the generalized degree of circular polarization $\hat{s}_{33}$, \textbf{(c,f,i)} the concurrence. }
    \label{fig:a0_1_FPW}
\end{figure}

If we investigate the nonlinear regime (in this case $a_0=1$), which can been seen in Fig. \ref{fig:a0_1_FPW}, the on- and off-shell contributions obtain ponderomotive broadening. The off-shell contribution is relatively less entangled compared to $a_0=0.1$. The on-shell contribution is still fully unentangled by the low recoil. The ponderomotive broadening combined with the relative decrease of the entanglement of the off-shell spectrum, does increase the difficulty of applying spectral filtering for higher intensity laser fields ($a_0 \gtrsim 1$).  

\onecolumngrid

\twocolumngrid

\bibliographystyle{apsrev4-1}
\bibliography{Bibliography.bib}
\onecolumngrid

\end{document}